\documentclass[
  12pt,
]{article}

\usepackage{arxiv}
\hypersetup{
    hidelinks              
}

\usepackage[round,authoryear]{natbib}
\usepackage{multibib}
\newcites{app}{Appendix References}

\title{
Estimating spatially-varying density and time-varying
demographics with open population spatial capture-recapture: a photo-ID
case study on bottlenose dolphins in Barataria Bay, Louisiana, USA}
\author{%
  Savannah Rogers\footnotemark[1] \\
  Centre for Research into Ecological and Environmental Modelling \\
  University of St Andrews, UK \\
  \AND
  Richard Glennie\footnotemark[1] \\
  Centre for Research into Ecological and Environmental Modelling \\
  University of St Andrews, UK \\
  \AND
  Len Thomas\footnotemark[2] \\
  Centre for Research into Ecological and Environmental Modelling \\
  University of St Andrews, UK \\
  \AND
  Todd Speakman \\
  National Marine Mammal Foundation \\
  Johns Island, SC, USA \\
  \AND
  Lance Garrison \\
  National Marine Fisheries Service \\
  Miami, FL, USA \\
  \AND
  Ryan Takeshita \\
  National Marine Mammal Foundation \\
  Johns Island, SC, USA \\
  \AND
  Paul van Dam-Bates \\
  Department of Fisheries and Oceans \\
  Nanaimo, BC, Canada \\
  \AND
  Lori Schwacke \\
  Marine Mammal Commission \\
  Bethesda, MD, USA
}
\date{September 16, 2026}

\begin{document}

\renewcommand{\thefootnote}{\fnsymbol{footnote}}
\maketitle
\footnotetext[1]{Joint first authors}
\footnotetext[2]{Corresponding author: len.thomas@st-andrews.ac.uk}
\renewcommand{\thefootnote}{\arabic{footnote}} 

\begin{abstract}
From long-term spatial capture-recapture (SCR) surveys, we can infer a
population's dynamics over time and distribution over space. It is
becoming more computationally feasible to fit these open population SCR
(openSCR) models to large datasets and build complex models with
spatially-varying density surfaces and time-varying population dynamics.
However, there is limited knowledge on how these methods perform when
applied to real data, especially in a maximum likelihood framework. We analyze a multi-year photo-identification survey of common bottlenose dolphins (\emph{Tursiops truncatus}) in
Barataria Bay, Louisiana, USA following the \emph{Deepwater Horizon} oil spill in \(2010\). Over
\(2000\) capture histories were collected between \(2010\) and
\(2019\). Using openSCR methods with non-Euclidean distances, we estimate time-varying population
dynamics and a spatially-varying density surface for this population. We
show that inference on survival, recruitment, and density over time
since the oil spill provides insight into increased mortality after the
spill, possible redistribution of the population thereafter, and
persistent low survival. Issues in the application are highlighted
throughout: possible model misspecification, sensitivity of parameters
to model selection, and difficulty in interpreting results due to model
assumptions and irregular surveying in time and space. For each issue,
we present practical solutions including assessing goodness-of-fit,
model-averaging, and clarifying the difference between quantitative
results and their qualitative interpretations. Overall, this case study
serves as a practical template other analysts can follow and extend; it
also highlights the need for further research on the applicability of
these methods as we demand richer inference from them.

\vspace{0.5cm}

\end{abstract}

\hypertarget{introduction}{%
\section{Introduction}\label{introduction}}

Capture-recapture studies collect repeated detections of identifiable
individuals and use these to infer how a population is changing over
time and space \citep{seber2019capture}. Open capture-recapture methods,
such as Jolly-Seber \citep{jolly1965explicit, schwarz1996general} and
Cormack-Jolly-Seber \citep{cormack1964estimates}, are the standard
methods applied to infer survival, recruitment, and abundance over time
\citep{williams2002analysis}. However, there is increasing use of open
population spatial capture-recapture (openSCR) methods
\citep{ergon_separating_2014, bischof_wildlife_2016, chandler_characterizing_2018, augustine_sex-specific_2020} which
incorporate spatial information on detection, conferring advantages such
as accounting for individual heterogeneity in detectability  due to spatial location,
reducing bias in survival estimates, and providing a rigorous estimation
of density \citep{gardner2010spatially, glennie2019open, efford2020spatial}.

Building complex models with spatially-varying density and time-varying
population dynamics is now feasible. Nevertheless, analyses attempting to combine these model components together and
apply them to a real dataset remain scarce \citep[e.g.,][]{bischof_estimating_2020}. No analyses combining time-varying population dynamics and spatially-varying density is known to these authors to exist in a maximum likelihood estimation (MLE) framework (although note \citet{winton_open_2023} includes time-varying population dynamics). Discussion of how this can be achieved is
important: openSCR poses new challenges, requires novel modeling
decisions, and necesitates different interpretations compared to non-spatial
analyses or non-openSCR analyses, e.g., in delimiting the
spatial domain of integration \citep{gardner2018state}, interpreting
recruitment/survival \citep{glennie2019open, ergon2014separating}, and
assessing goodness-of-fit.

In this paper, we build upon the methods of \citet{glennie2019open} to meet the needs of our case study: a MLE framework open SCR analysis of a
multi-year, boat-based photo-identification (photo-ID) capture-recapture
survey on common bottlenose dolphins (\emph{Tursiops truncatus},
hereafter referred to as dolphins) in Barataria Bay, Louisiana, USA
\citep{mcdonald2017survival}. We describe each important step in the
analysis workflow: (1) formatting photo-ID data into a form amenable to
such analyses; (2) fitting complex open SCR models in novel combinations
including spatially-varying density and time-varying population
dynamics; (3) drawing appropriate quantitative and qualitative
inference; and (4) assessing goodness-of-fit for the model.

The dolphin population in Barataria Bay was affected by heavy oiling from the \emph{Deepwater Horizon}
oil spill in April 2010 \citep{hayes2018us}. As part of the \emph{Deepwater Horizon} (DWH) Natural
Resource Damage Assessment (NRDA), boat-based photo-ID surveys were
conducted post-spill between \(2010\) and \(2014\). An additional survey
was conducted in \(2019\) to inform an Environmental Impact Statement
required for a planned DWH restoration project in Barataria Bay \citep[the
Mid-Barataria Sediment Diversion,][]{garrison2020predicting}.
\citet{mcdonald2017survival} fit a Bayesian openSCR model to data
available up to \(2014\) to infer density over four spatial strata and
estimated time-varying survival; however, the computational burden was
overwhelming (\(\sim 2000\) MCMC iterations took \(9\) days) and so
adding further model complexity was prohibitive. Here, the goal of the
analysis is to take advantage of recent computational efficiencies accomplished with MLE \citep[see benchmark comparison in][]{glennie2019open} to
incorporate data collected in \(2019\), which also includes an area not
surveyed prior to \(2019\), and to fit a wider range of population
dynamics and density models, allowing for us to select between them.
This will provide richer and updated inference on a population whose
assessment contributes to understanding the long-term effects of the oil
spill \citep{schwacke2017quantifying, takeshita2017deepwater}.

This case study shows the advantages these more complex methods can have
for assessing populations. The problems that arise in the application
are also likely to be common with similar studies: sensitivity of
inference to model selection, aspects of poor goodness-of-fit due to
current modeling limitations, and the need for cautious interpretation
of inference due to model assumptions or spatio-temporal irregularity in
sampling. For those who apply these methods in similar studies, this is meant to inform them of the method's performance given current
modeling capabilities; for researchers in openSCR, we hope this will
provide grounds for future method development.

\hypertarget{materials}{%
\section{Data and Methods}\label{materials}}

In this section we describe the important steps of the analysis:
formatting the data and covariates (Section \ref{data}) and 
specifying the components of the model (Section \ref{model}).
We assume the reader is familiar with standard SCR
methods \citep[e.g.,][]{borchers2008spatially, royle2013spatial}.

\hypertarget{data}{%
\subsection{Data}\label{data}}

\begin{figure}
\includegraphics[width=0.9\textwidth]{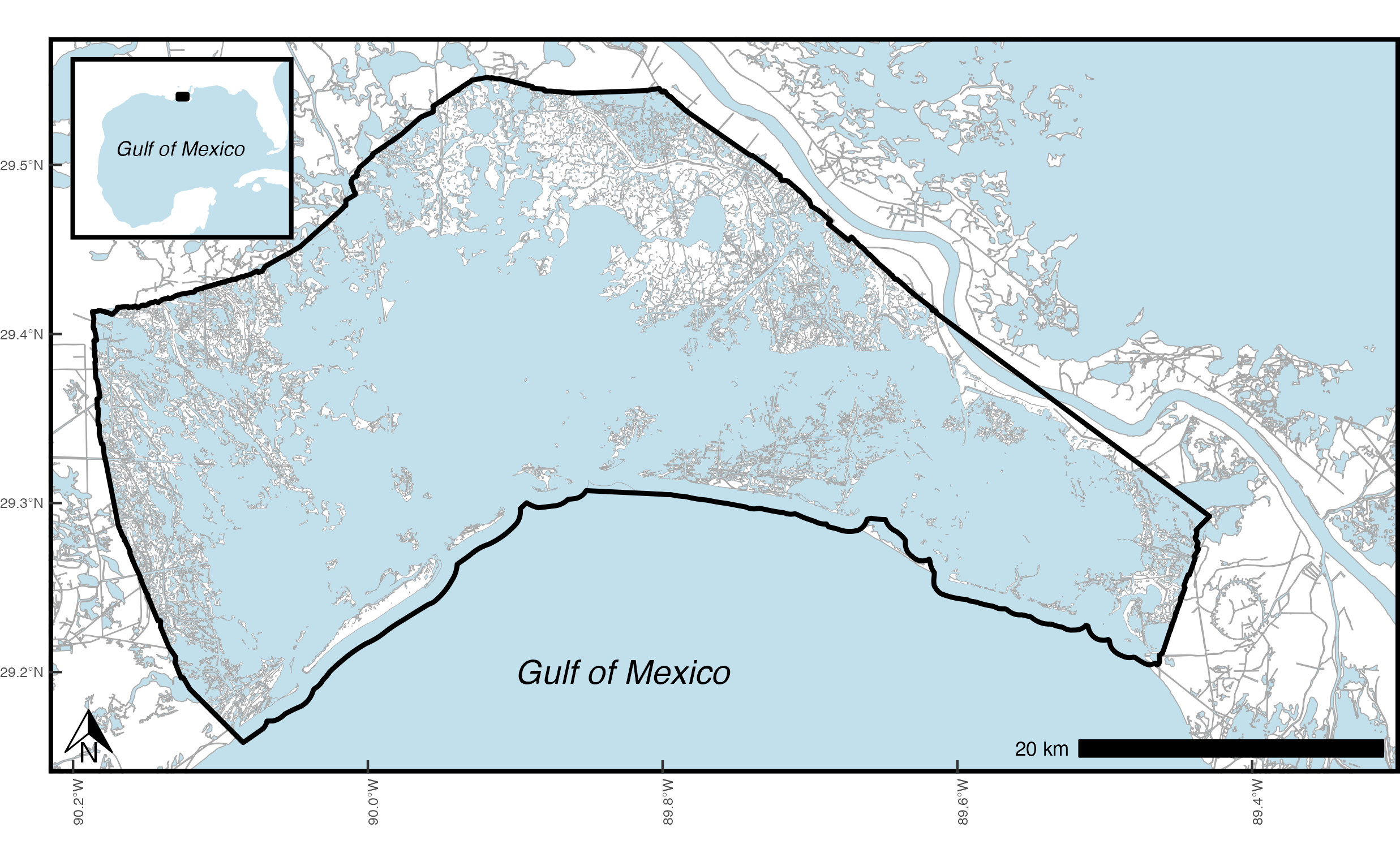} \caption{Barataria Bay study area. Inset map shows location of Barataria Bay (black square) within the Gulf of Mexico. Basemap sources: NOAA.}\label{fig:study_area}
\end{figure}

The defined study area within Barataria Bay (Figure
\ref{fig:study_area}) is approximately 1280~km$^2$ and is divided into four strata: west (406~km$^2$), central (285~km$^2$),
southeast (496~km$^2$), and islands (93~km$^2$).  A detailed capture history summary is provided in Appendix 1. The survey protocol has been previously described
in \citet{mcdonald2017survival}. A notable addition to the previous
study is the extended survey effort in \(2019\) of the southeast stratum
as, prior to \(2019\) only the island, west, and central strata were
surveyed \citep{garrison2020predicting}. This addition to the study area
(in space and time) has important ramifications for inference.

Survey data on individuals consist of photographs of dolphin groups
sighted during on-effort surveying. We only include sightings of distinct individuals in our analysis (i.e., ``very distinctive'' or ``average distinctive'' according to \cite{urian_recommendations_2015} as adapted for dolphins from \cite{friday_measurement_2000}, \cite{urian_status_1999}). Distinctive marks are acquired
randomly (e.g., tooth rakes and scars from entanglements with fishing gear); we do not go into detail here on this process (see \citet{melancon2011photo};
\citet{mcdonald2017survival}), but a key
observation is that not all individuals in the population are distinctive and while an individual may become more distinctive, they can still be reliably identified and their distinctiveness can
only improve. We found the proportion of distinctive individuals in our data to vary across space but not time (see Appendix 2), so we scale spatial density estimates by the predicted proportion distinctive at each location. Therefore, reported
abundance and density is for the whole population.

\hypertarget{traps}{%
\subsubsection{Occasions and traps}\label{occandtraps}}

The application of SCR to this photo-ID survey differs from standard
applications as the detector, a boat in this case, moves in continuous
space. Therefore, within the SCR framework, the encounter rate between
an individual animal and the detector depends on the path the detector
has taken with respect to that individual's activity center. This can be
accounted for within the standard open population SCR model by redefining
what is meant by a ``trap''.

The detector data consists of GPS records, approximately every
\(30\) seconds, of the boat's location whilst surveying. Here, a survey is defined as a single trackline traversed by the boat. These surveys
were grouped into a robust design \citep{pollock1982capture}: secondary
occasions comprised between 2--11 surveys and primary occasions contained
3--4 secondary occasions. Secondary duration was 2--3 days. Primaries spanned 7 to 17 days with 50 to 1777 days between primaries. Start dates and primary-specific timespans are given in Appendix 3. With respect to an openSCR analysis, the
construction of secondary and primary occasions is necessary as it is
assumed the first sighting of an individual within each secondary occasion is independent of first sightings in other secondary occasions (note that we only retain for analysis the first detection of an individual within each secondary occasion to avoid spatial autocorrelation between subsequent sightings) and
demographic states do not change within primary occasions. 

Within each secondary occasion, we must be able to quantify the
probability an individual, given their activity center, will be
detected. As alluded to above, this depends on the continuous path the
detector traveled. Following search-encounter methods for SCR
\citep{royle2011spatial}, we approximate this continuous process by
segmenting the boat's path, linearly interpolating between GPS
locations. We then construct a \(1\)km grid across the survey area and
record how many surveys traversed each grid cell. We term this quantity
\emph{effort}. ``Traps'' are thus defined to be located at the center of grid cells
that have non-zero effort, and sightings were allocated to the trap associated with the grid cell
within which the sighting took place.

\hypertarget{mesh}{%
\subsubsection{Mesh}\label{mesh}}

In SCR, each individual is associated with a location in space: its
activity center (AC). This location is unknown and so we must average over
all possible locations of that AC. To do this we must
specify a set of locations within which each individual's AC must reside. The extent of this spatial grid, termed a
\emph{mesh} or \emph{mask}, is application-specific. In this case, the
mesh was constructed from a grid of \(1\)km grid cells within an area
delineated by the bay's barrier islands and by analysis of dolphin
telemetry data collected within the bay \citep{wells2017ranging}. We excluded any mesh location grid cells centered on land to prevent ACs from occurring on land. While SCR typically requires a mesh buffer of $3-4 \sigma$, this is not the case when ACs cannot occur outside of a bounded area (see Appendix 4). Telemetry data from \cite{wells2017ranging} shows that dolphins in Barataria Bay exhibit high site fidelity and have utilization distributions within the bay: the average maximum distance from shore for dolphins that primarily occupied the area around the west barrier islands was only $1.75$km. We therefore
extended the mesh beyond the barrier islands by \(1\)
kilometer, allowing for the possibility that ACs may
marginally lie outside the delineated area.  

Three spatial covariates were used in this analysis (details provided in Appendix 5). The stratum was
defined based on prior evidence of differences in dolphin movement
between these areas \citep{wells2017ranging}. Openness was also defined based on existing evidence of variation: the
bay is a highly heterogeneous habitat, especially in the north where the
definition of land and water is itself changing over time. In
exploratory analysis, we found that empirical encounter rates were
notably higher in areas that were more sheltered from poor weather and
in order to capture this variation, we delineated open and sheltered areas based on the experience of our field crew in the area. Finally, the averaged salinity covariate represents the mean salinity within each grid cell from Jan 2011 to Dec 2017 (all available data, see Appendix 6) based
on hydrodynamic modeling \citep{hornsby2017using}.
Thus, this salinity covariate is averaged over both space and time and
can only reflect the long-term relationship between the activity
center density and average salinity level. Salinity levels change rapidly within
the bay and so more fine scale modeling would be necessary to capture the
relationship between where individuals go each day and the salinity
level on that day. Plots of each covariate are included in the
supplementary material (Appendix \(5\)).

\hypertarget{model}{%
\subsection{Model}\label{model}}

The MLE open population SCR modeling framework is described in
\citet{glennie2019open}. Three model components must be specified:

\begin{itemize}
\item
  \textbf{Detection model}: the probability a sighting occurs within
  each secondary occasion given an individual's AC (Section
  \ref{detection});
\item
  \textbf{Population dynamics model}: the temporal process over primary
  occasions controlling when individuals are available and unavailable
  for detection (Section \ref{popdyn});
\item
  \textbf{Density model}: the point process model that determines where
  ACs arise in space (Section \ref{density}).
\end{itemize}

\hypertarget{detection}{%
\subsubsection{Detection}\label{detection}}
In a study system where areas of habitat may be bounded by non-habitat that animals cannot cross, it is important to consider what this means for the detection function. For our system, the Euclidean distance between a dolphin's AC and water on the other side of an island may be relatively small. However, since the dolphin cannot travel through the island, the functional distance from its AC and the water on the other side is much longer than the Euclidean distance.  We therefore used a non-Euclidean distance (first proposed in SCR by \citet{royle_spatial_2013}) ``as the dolphin swims'' by calculating the shortest distance path through water between an animal's AC and a trap. This was computationally efficient as the distance between each mesh point and trap could be calculated once, before model fitting, using Dijkstra's algorithm \citep{gdistance}.  \\
Given an individual's AC \(\bm{x}\), the first encounter
rate with trap \(j\) on primary occasion \(k\) in secondary occasion
\(l\) is given by
\(e_{j,k,l}(\bm{x}) = \lambda_{j,k,l}\exp\{-d_j(\bm{x})^2/(2\sigma_{j,k,l}^2)\}\)
for base encounter rate parameter \(\lambda_{j,k,l}\) and encounter
range \(\sigma_{j,k,l}\) and where \(d_j(\bm{x})\) is the non-Euclidean
distance between trap \(j\) and AC \(\bm{x}\). This formulation is the
well-known hazard half-normal detection function
\citep{borchers2008spatially}.

The probability of an individual being sighted on one of the \(J\) traps
is therefore
\(p_{k,l}(\bm{x}) = 1 - \exp\left\{-E_{k,l}(\bm{x})\right\}\) where
\(E_{k,l}(\bm{x}) = \sum_{j = 1}^J e_{j,k,l}(\bm{x})u_{j,k,l}\) and
\(u_{j,k,l}\) is the effort for trap \(j\) on primary occasion \(k\),
secondary occasion \(l\). Given an individual is seen by one of the
traps, the probability the sighting occurred on trap \(j\) is given by
the relative encounter rate
\(e_{j,k,l}(\bm{x})u_{j,k,l} / E_{k,l}(\bm{x})\).

Overall, the probability of a first sighting on trap \(j\) in primary
occasion \(k\), secondary \(l\) of an individual with AC
\(\bm{x}\) is given by
\(p_{k,l}(\bm{x})e_{j,k,l}(\bm{x})u_{j,k,l} / E_{k,l}(\bm{x})\). This is
known as the competing hazards detection model
\citep{borchers2008spatially}. We note that for our model, encounter rate parameters $\lambda_{j,k}$ and $\sigma_{j,k}$ vary with trap and primary, but do not vary across secondaries within a primary (see Appendix 7). We leave the above notation to clarify the general case, but specify the encounter rate hereafter as $e_{j,k}$.

\hypertarget{popdyn}{%
\subsubsection{Population dynamics}\label{popdyn}}

Whether an individual is available for detection or not during a primary
occasion depends on its \emph{state}. In \citet{glennie2019open},
individuals have three states (``unborn'', ``alive'', and ``dead'') and
are only available for detection in one of them (``alive''). Individuals
change states between primary occasions and not within (see closure
assumption discussed above).

It is important to remember that these states are inferred from an
individual's detection availability and not from observed life events.
Conceptually, each individual's encounter history is explained by a
possible period of no sightings (while ``unborn'' they are unavailable for
detection), a mix of sightings and no sightings coherent with the
detection process (``alive'') and then possibly a period of no further
sightings (``dead'' individuals are no longer available).

We will enforce that individuals must begin in primary
occasion \(1\) in either state ``unborn'' or ``alive'' and then proceed through
the states in the sequence ``unborn'' \(\rightarrow \) ``alive'' \(\rightarrow \)``dead'' until all
individuals reach state ``alive'' or ``dead'' by the end of the survey.

This state process is defined by analogy to
population dynamics: individuals enter a population by immigration or
being born, live for some period where they may be detected, and then
emigrate or die. It is through this analogy that the probability of the
transition ``unborn'' \(\rightarrow\)``alive'' is called recruitment probability and
``alive'' \(\rightarrow \)``dead'' survival probability. 
Because we record detections of distinctive individuals, the ``alive'' state refers to animals that are alive and distinctive, and the ``unborn'' state may contain animals in the population that have not yet become distinctive. The process of recruitment is therefore a combination of entering the population and acquiring distinctiveness, and survival therefore pertains to survival of distinctive individuals. For survival probability it is
well known that estimates can be biased by unmodeled heterogeneity or
emigration
\citep{efford2020spatial, ergon2014separating, kendall1997estimating}
and is termed \emph{apparent} survival. A similar argument applies to
recruitment.

For this application, we define both the survival and recruitment
processes in continuous time. Let \(1-\phi_k^{\delta_k}\) be the
probability of the transition ``alive'' \(\rightarrow \)``dead'' between primary
occasions \(k\) an \(k + 1\) where \(\delta_k\) is the time (in years) between the
mid-points of both occasions.  Therefore $\phi_k$ is the annualized average survival probability between primaries $k$ and $k+1$.

For transitions ``unborn'' \(\rightarrow\)``alive'', let \(\gamma_k\) be the transition
rate between primary occasions \(k\) and \(k + 1\) and \(\beta_k\) the
probability of the transition between these primaries. We define
\(\beta_1 = \exp\left(-\sum_{k = 1}^{K - 1} \gamma_k\delta_k\right)\)
and
\(\beta_k = (1 - \beta_1)\gamma_k\delta_k / \sum_{m = 1}^{K - 1} \gamma_m\delta_m\)
for primary occasion \(k > 1\).

\hypertarget{density}{%
\subsubsection{Density and Abundance}\label{density}}

To obtain estimates of density and abundance of alive individuals for the entire population (both distinctive and non-distinctive), we use the density of ACs for distinctive individuals, population dynamics, and proportion distinctive models. We do not directly model the AC density of distinctive individuals in the population at any particular time; rather we model the AC density of distinctive
individuals \emph{alive at some point during the survey}. This is a
point pattern in space and does not depend on time. We assume ACs arise from an inhomogeneous Poisson process with intensity
\(D(\bm{x})\) at location \(\bm{x}\) \citep{borchers2008spatially}.
Defining \(\overline{D}\) as the mean value of \(D\) over the study
region, this implies that the probability density of an AC
occurring at location \(\bm{x}\) is \(D(\bm{x}) / \overline{D}\) \citep{diggle2013statistical}.

Expected alive, distinctive individual AC density for any given time and location is derived jointly
from the population dynamics and AC density models. Defining
\(D_k(\bm{x})\) to be the alive, distinctive individual AC density during primary occasion \(k\) at
location \(\bm{x}\), we have that \(D_1(\bm{x}) = \beta_1 D(\bm{x})\)
and that for \(k > 1\),
\(D_k(\bm{x}) = \phi_{k - 1}^{\delta_{k - 1}} D_{k - 1}(\bm{x}) + \beta_k D(\bm{x})\). In other words, for the first primary, the density of alive distinctive individuals at $x$ is the density at $x$ of distinctive individuals alive at some point during the study ($D(x)$) times the probability that an individual recruits before the first primary ($\beta_1$). For subsequent primaries, the density of alive distinctive individuals at $x$ is the density of distinctive individuals alive at $x$ in the previous primary ($D_{k-1}(x)$) times the probability that those animals survived since the previous primary ($\phi_{k-1}^{\delta_{k-1}}$) plus the number of distinctive individuals that recruited between the previous and current primary at $x$ ($ \beta_kD(x)$). This means that while the relative shape of the expected distinctive individual AC density surface remains consistent through time, the intensity of that surface, and the resulting impact on abundance, changes by primary according to the population dynamics model. Because the density of ACs alive at some point during the survey is not biologically informative, we calculated the average number of distinctive animals alive at any point during the survey according to the population dynamics model to derive the mean expected AC density of alive, distinctive individuals. 

We also used the expected AC density surface of alive, distinctive individuals to generate an alive, distinctive ``animal density surface'' which we denote $D^\alpha$. While the expected AC density surface is what has historically been presented in SCR analyses \citep{durbach_thats_2024}, it only reflects the density of ACs rather than how animals use space \citep{royle_integrating_2013}. Since animals move around their AC, we sought to better display this information by defining the alive, distinctive animal density at $x$ in primary $k$ as $D^\alpha_{k}(x) = \sum_{y \exists \bm{M}} \frac{e_{x,k}(y)}{\sum_{z \exists \bm{M}} e_{x,k}(z)} * D_k(y)$ where $\bm{M}$ are all locations in the mesh, $D_k(y)$ is the alive, distinctive AC density at $y$ in primary $k$, and $e_{x,k}(y)$ is the encounter rate as defined above for a trap located at $x$ and an individual with AC $y$. This more intuitively reflects higher densities in areas where animals are seen more often but still reflects the highest density at the ACs. 

Because the proportion distinctive is estimated based on where animals are seen (see Appendix 2), we used it to scale the alive, distinctive animal density surface to estimate alive animal density for the entire population. The total alive animal density is defined $D_{k}^{\alpha*}(x)=\frac{D_{k}^\alpha(x)}{r(x)}$ where $r(x)$ is the estimated proportion of distinctive individuals at $x$. We therefore estimated total alive animal abundance for primary $k$ as $N_k = \frac{A}{n_M}\sum_{x\exists \bm{M}}D^{\alpha*}_k(x)$ where $A$ is the total area of the mesh and $n_M$ is the number of mesh cells.

\hypertarget{fitting}{%
\subsubsection{Fitting the model}\label{fitting}}

We fit all models in the \texttt{R} package \texttt{openpopscr}
(available at \url{https://github.com/SavannahAlicia/openpopscr}). Each
parameter could depend on covariates, either linearly or through thin
plate regression splines \citep{wood2003thin} with fixed numbers of basis functions (hyper-parameter $k$, see Table \ref{tab:models} for all considered parameter specifications) and no penalty term. We restricted attention to models where \(\lambda, \sigma\)
could depend on stratum, openness, and primary; \(\phi, \beta\) could
depend on time; and \(D\) could depend on spatial location and average
salinity.

\hypertarget{select}{%
\subsubsection{Model Selection}\label{select}}

Models (and covariate effects) were selected using Akaike's information
criterion \citep[AIC; ][]{akaike1998information}. Due to computational
burden, regression splines were limited to a number of basis functions (hyper-parameter $k$) with a
maximum of \(10\) for 1D smooths and \(20\) for 2D. Uncertainty in hyper-parameter
$k$ for splines was incorporated by retaining what we refer to as the ``best model set'': the subset of all models with the same covariate structure aside from differing hyper-parameters that had AIC
scores less than two units from the model with lowest AIC score. This best model set was then used to produce model-averaged parameter estimates, as detailed below in Section \ref{inf}. 

\hypertarget{inf}{%
\subsubsection{Inference}\label{inf}}

Model-averaged parameter estimates and uncertainty in these estimates
were approximated by sampling with replacement models from the best model set with
probability proportional to their AIC weights and using a parametric
bootstrap of parameters from that model's maximum likelihood estimator's
asymptotic normal distribution. We term this a model-averaging
bootstrap, and we chose this approach to calculate uncertainty estimates for derived parameters.

\hypertarget{gof}{%
\subsubsection{Goodness-of-fit}\label{gof}}

Goodness-of-fit was assessed by comparing data simulated given our
inferred parameters with the observed data. To compare the simulated
data with the observed, we conducted three randomization tests
\citep{manly2006randomization}: (1) Recruitment: the number of
individuals seen for the first time on each primary occasion; (2)
Survival: the average number of primaries between the first and last
sighting of an individual; (3) Density: the number of individuals
detected on each trap over the entire survey period.

\hypertarget{results}{%
\section{Results}\label{results}}

Surveys were collected into \(34\) secondary occasions comprising on
average \(5\) surveys. These were then organized into \(11\) primary
occasions. Duration of inter-primary intervals was irregular, being
approximately \(0.5\) years, except for the last primary occasion (11)
which occurred \(5\) years after primary occasion \(10\) (Supplementary Table 2). 

\begin{figure}
\includegraphics{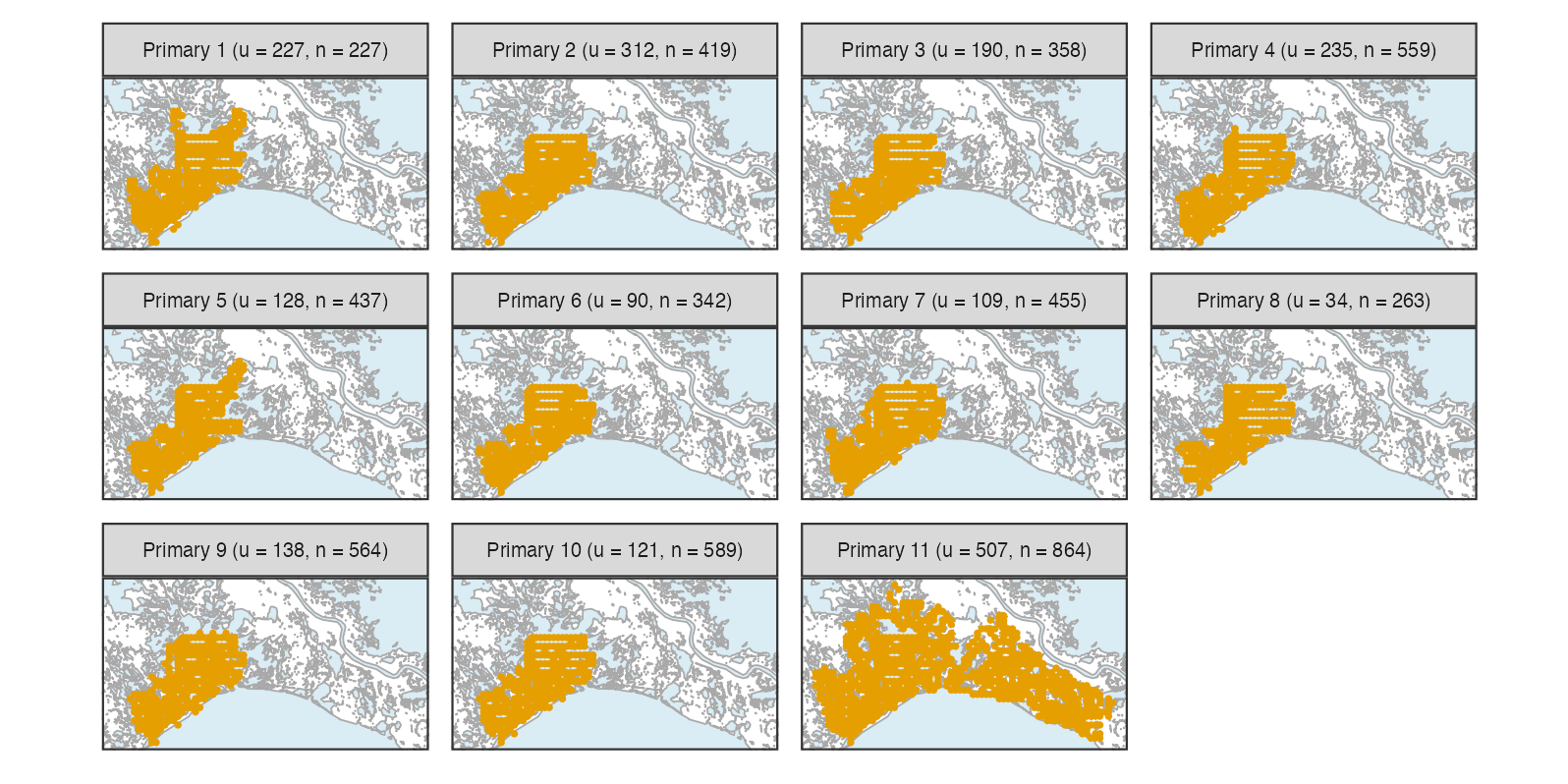} \caption{Location of surveyed traps (yellow points) for each primary occasion.  Panel headings show, in brackets, number of individuals seen for first time $(u)$ and total number of individuals seen $(n)$ on that primary occasion.}\label{fig:effort}
\end{figure}

In total, 2091 unique individuals were sighted during the entire survey
period over a constructed grid of \(909\) traps and  a
constructed mesh of 1280 points. Figure \ref{fig:effort} shows the traps surveyed, the number of
individuals sighted for the first time (\(u\)) in each primary occasion, and the total number of individuals seen (\emph{n}).
The inference drawn from the analysis must be considered in light of the
spatial and temporal irregularity in the data. Many individuals were seen for the
first time in primary \(11\) (and most in the southeast) and so there is
no new temporal information on the survival rates of these most recent
individuals; the inferences made on survival stem from recaptures in
primary occasion \(11\) of those individuals seen in primaries surveyed
five years earlier, which covered a subset of the study area.

\hypertarget{model-selection}{%
\subsection{Model Selection}\label{model-selection}}

We considered many possible parameterizations (Table \ref{tab:models}) for the model and evaluated them by AIC to choose the best model. 

\begin{table}[ht]
\centering

\begin{tabular}{p{2.6cm}p{2.6cm}p{2.4cm}p{2.4cm}p{3.5cm}}
  \hline
Detectability \newline ($\lambda_0$) & Movement \newline ($\sigma$) & Recruitment ($\beta$) & Survival \newline ($\phi$) &  Density  \\ 
  \hline
constant & constant & constant & constant & constant \\ 
  trap stratum & trap stratum & s(time, \newline $k = 3:7$) & s(time, \newline $k = 3:7$) & s(2 dimensional space, $k = 5:20$) \\ 
  trap stratum+ \newline trap openness & trap stratum+ \newline trap openness & factor(primary) & factor(primary) & s(2 dimensional space, $k = 5:20$ )+ \newline s(salinity, \newline $k = 3:7$) \\ 
    trap stratum+ \newline trap openness+ \newline factor(primary) & trap stratum+ \newline trap openness+ \newline factor(primary) &  &  &  \\ 
   \hline
\end{tabular}
\caption{All possible model parameterizations considered. Each column represents a parameter, and each row lists a set of covariates considered for that parameter.
Splines (denoted with an ``s'') are given with the range of hyperparameters $k$.  All combinations of parameterizations for each parameter were considered.
See Appendix 5 for descriptions of covariates.} 
\label{tab:models}
\end{table} 
Model selection using AIC was successful in discriminating between models with differing  factor covariate effects (e.g.,~stratum) on parameters. However, AIC
discriminated poorly between models with different numbers of basis functions in the
regression splines (hyper-parameter $k$, Supplementary Table 4). For survival and recruitment, there was
clear support for temporal variation, but uncertainty when using AIC to
select how smooth the temporal relationships were (Supplementary Table 4).
For density, AIC failed to identify an optimal smoothness below the
computational maximum limit set.

The best model set had the form:

\[
\begin{aligned}
\lambda &\sim \text{stratum} + \text{openness} + \text{primary occasion} \\
\sigma &\sim \text{stratum} + \text{openness} + \text{primary occasion} \\
\gamma &\sim s(\text{time}, \text{df} = a) \\
\phi &\sim s(\text{time}, \text{df} = b) \\
D &\sim s(x, y, \text{df} = 20) + s(\text{averaged salinity}, \text{df} = 5) 
\end{aligned}
\]

where \(\sim\) denotes that the working parameter corresponding to the
parameter on the left-hand side had covariate effects for those
covariates on the right-hand side. The \(s\) symbol indicates a
regression spline was used and \(\text{df}\) gives its degrees of
freedom. The \(7\) best AIC models had hyper-parameters \((a,b)\) in
the set \(\{(5,5), (6,3), (6,4), (6,5), (7,3), (7,4), (7,5)\}\).

We now review the results obtained from \(10000\) simulations from the
model-averaging bootstrap (Section \ref{inf}) of these six models.
Results for detection parameters (encounter rate $\lambda$ and encounter range $\sigma$) are included in Appendix \(7\).

\hypertarget{surv}{%
\subsection{Survival}\label{surv}}

\begin{figure}
\includegraphics[width=0.8\textwidth]{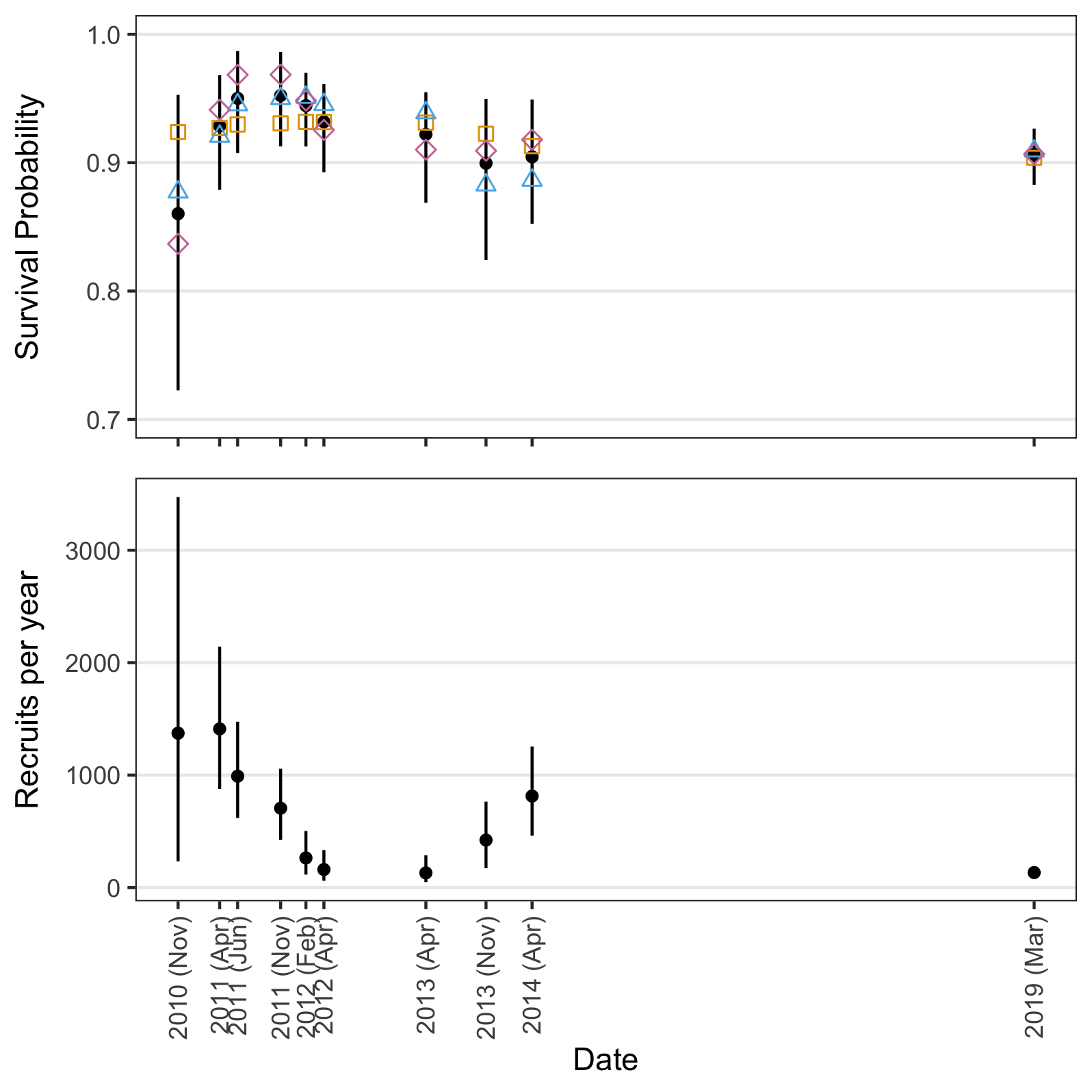} \caption{Mean annual survival probability estimates ($\phi$) for the preceding inter-primary period with mean annual survival probability estimates from contributing models with three (squares), four (triangles), and five (diamonds) basis functions (upper panel); mean estimated number of recruits per year between primary occasions (values on the x-axis correspond with the primary of arrival, bottom panel). Vertical black lines in both panels display approximate $95\%$ confidence interval over primary occasion from $10000$ model-averaging bootstrap samples.}\label{fig:survplot}
\end{figure}

We refer to average annualized survival probability between primary period $k$ and $k+1$ (i.e., $\hat{\phi}_k$) as annual survival probability for inter-primary interval $k$.  
Estimated annual survival probability (Figure \ref{fig:survplot}) initially
increases from inter-primary interval \(1\) until inter-primary interval \(3\) and
then more gradually decreases from inter-primary interval \(4\) until inter-primary interval \(10\). Discrepancy
between the different possible temporal regression splines on survival
is largest around the initial increase in survival with more smooth
models predicting a less rapid increase compared to those with more basis functions.

\hypertarget{recruitment}{%
\subsection{Recruitment}\label{recruitment}}

Recruitment had a consistent pattern across models in the best model set.
Recruitment is relatively high immediately after the oil spill (Figure
\ref{fig:survplot}) and then steadily decreases prior to \(2014\) when a
short increase in recruitment occurs. Recruitment is then estimated to
be relatively low in the five years prior to the \(2019\) surveys.

\hypertarget{resdensity}{%
\subsection{Density and Abundance }\label{resdensity}}

\begin{figure}
\includegraphics[width=0.8\textwidth]{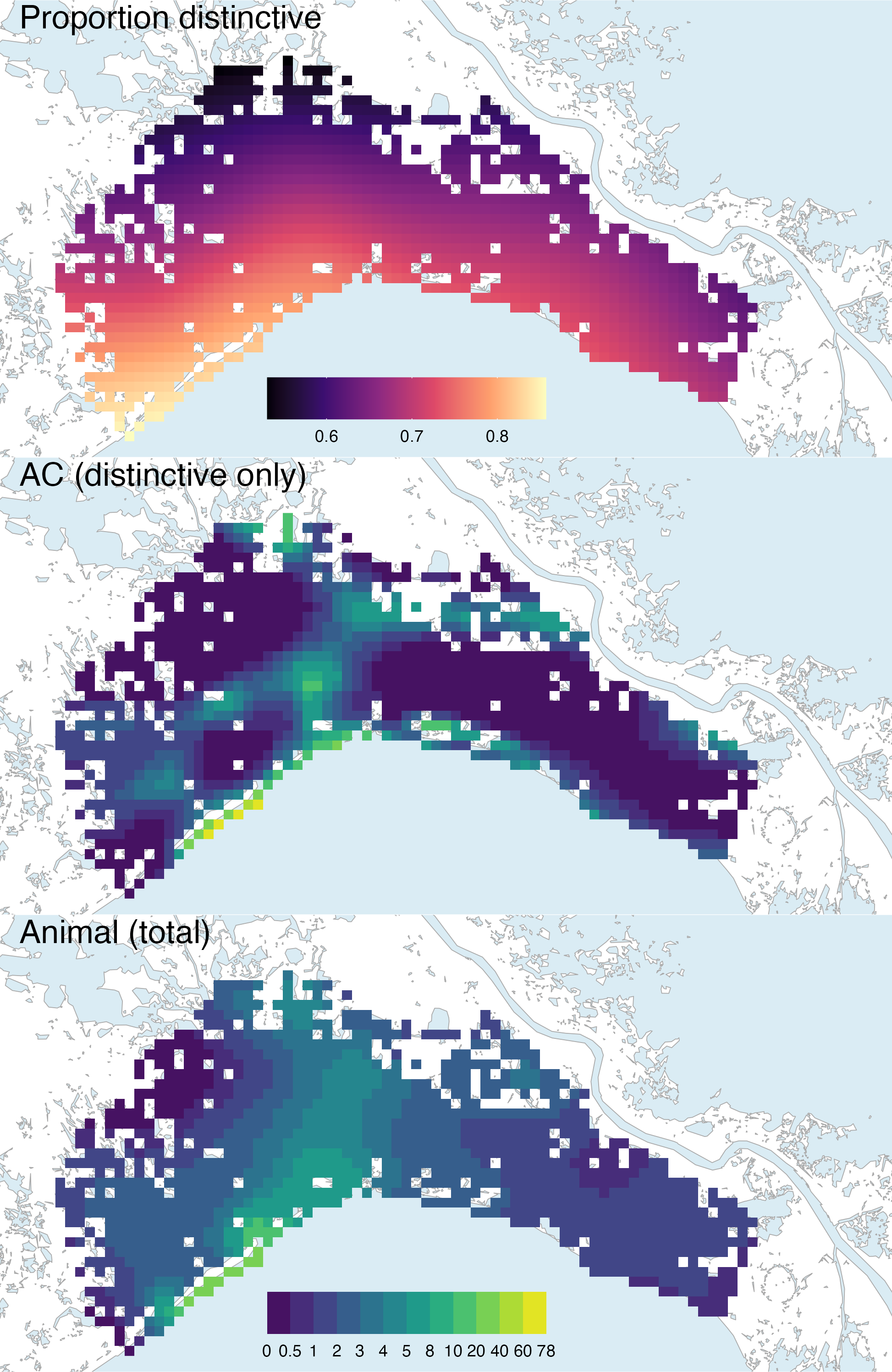} \caption{Estimated proportion distinctive animals (top panel); mean expected activity center (AC) density of alive, distinctive individuals ($D$, AC per km$^2$, middle panel); derived mean alive animal density for the total population ($D^{\alpha*}$, animals per km$^2$, bottom panel). Note that color bins for densities are unequal sizes. AC and animal densities are from $10000$ model-averaging bootstrap samples.}\label{fig:meanD_nocut}
\end{figure}

The proportion of distinctive individuals varied across space, with consistently higher values where the Bay meets the Gulf and the highest proportion southwest of the barrier islands (Figure \ref{fig:meanD_nocut}, top panel). The expected alive, distinctive AC density and estimated mean total alive animal density are heterogeneous across the study area with
high density around the islands and passes between the Bay and the Gulf
(Figure \ref{fig:meanD_nocut}, middle and bottom panels, respectively). Total alive animal density is lowest in the northwestern, southeastern, and northern regions bounding the study area. Overall, mean total alive animal density is
lower in the southeastern region of the study area compared to the west
and central areas.
In the south, density appears to increase close to the islands. Mean density for the entire $1280$km$^2$ study area was 2.70 dolphins per total km$^2$, or 3.05 dolphins per water km$^2$.

\begin{figure}
\includegraphics{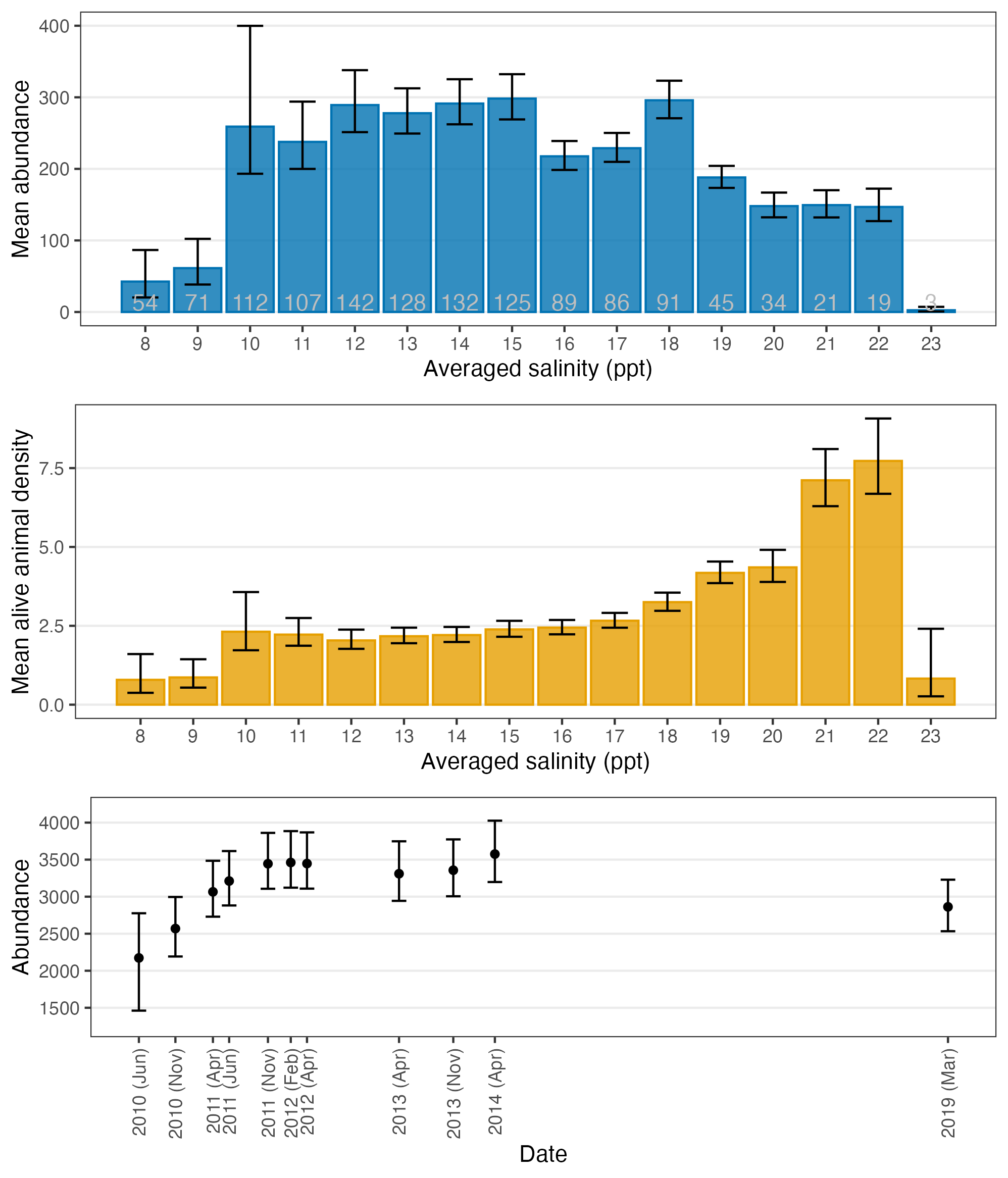} \caption{Estimated abundance (upper panel), and density (alive animals in the total population per square kilometer, $D^{\alpha*}$, center panel) averaged over all primaries in each salinity (ppt) band (where a salinity band with label $x$ includes all mesh points with salinities in the interval $[x - 0.5, x + 0.5)$). Numbers at base of each bar in top panel is the area (km$^2$) covered by each salinity band. Estimated mean total abundance for each primary occasion ($N_k$, lower panel) with approximate $95\%$ confidence interval (vertical lines). All from $10000$ model-averaging bootstrap samples with approximate $95\%$ confidence intervals.}\label{fig:salplot}
\end{figure}

Density was also a function of averaged salinity. We interpret the
salinity effect using the mean alive animal density and abundance for the total population within (binned) salinity bands (Figure
\ref{fig:salplot}).

Abundance within the region of inference initially increases from
\(\sim 2173\) in primary occasion \(1\) (Jun \(2010\)) to \(\sim 3445\)
in primary occasion \(5\) (Nov \(2011\)), then varies
around \(\sim 3500\) up to primary occasion \(10\) (Apr \(2014\)),
before declining to \(\sim 2862\) in primary occasion
\(11\) (Mar \(2019\)) (Figure \ref{fig:salplot}).

\hypertarget{resgod}{%
\subsection{Goodness-of-fit}\label{resgod}}

\begin{figure}
\includegraphics{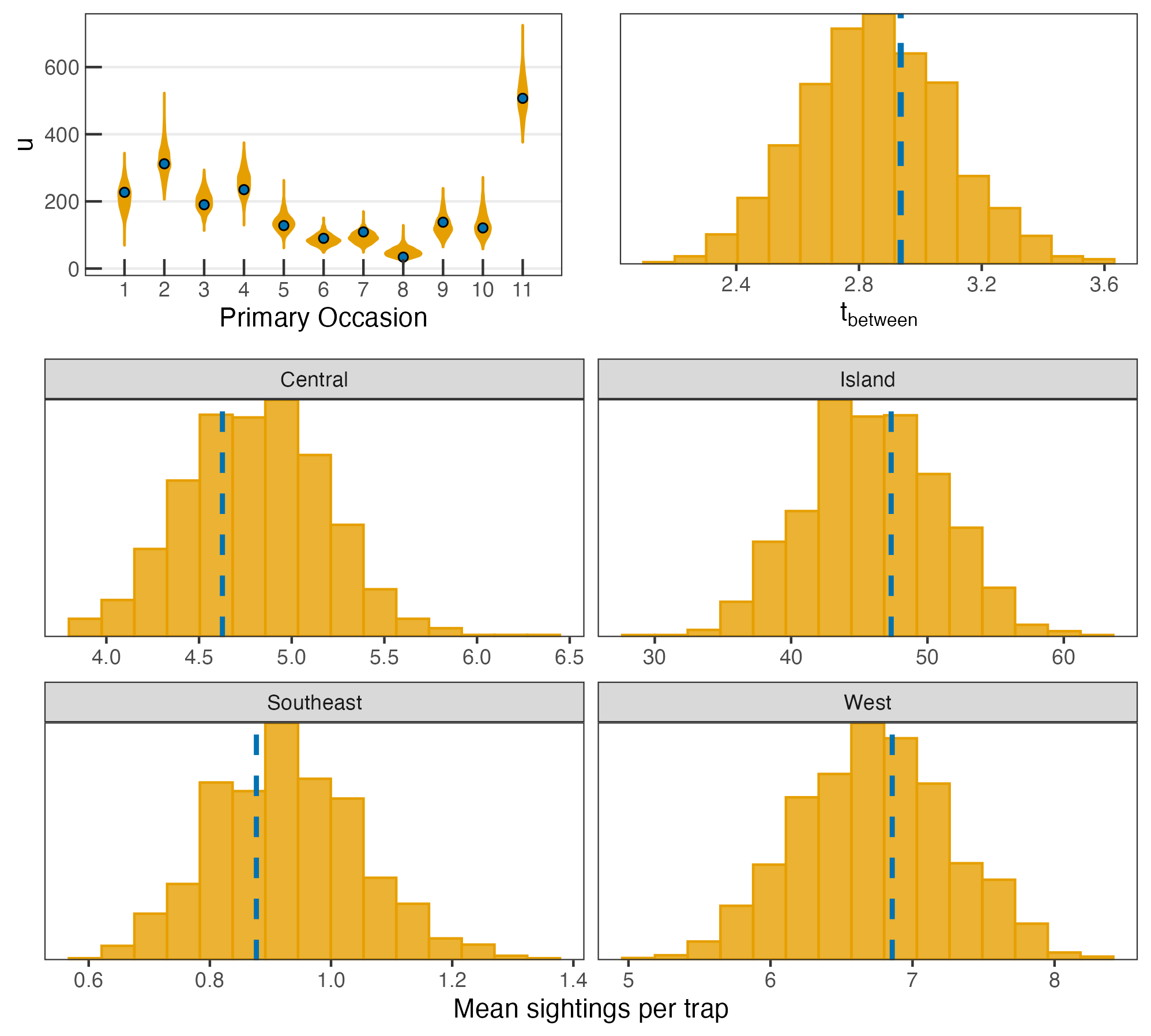} \caption{Goodness-of-fit tests: (top-left) distribution of the simulated number of individuals seen for the first time (u) for each primary occasion (blue violins); (top-right) distribution of mean time between first and last sighting of an individual; (bottom) distribution of mean number of individuals sighted per trap in each spatial stratum. All distributions are derived from $10000$ simulated datasets; solid points and dashed lines give the corresponding statistic computed from the observed data.}\label{fig:uplot}
\end{figure}

Goodness-of-fit tests showed evidence of general good fit to multiple
aspects of the data (Figure \ref{fig:uplot}). 
The expected number of new individuals sighted, \(u\), for each primary
occasion was well described for most primary occasions, except for primary
occasions \(7\) and \(8\) where the model very slightly over- and under-predicted
recruitment respectively, indicating a lack of ability in the model to
capture extreme changes in recruitment that both of these primaries may
have experienced.  The observed and expected mean time between first and last sighting of
an individual (\(\text{t}_{\text{between}}\)) gave no evidence of poor
model fit.  The number of individuals sighted on each trap within each stratum
indicated adequate model fit for all stratum.

\hypertarget{discussion}{%
\section{Discussion}\label{discussion}}

In this final section, we interpret the results of our case study in the context of model assumptions (Section \ref{casestudy}), discuss the advantages and limitations of this analysis (Section \ref{limitations}), and present
opportunities for future research (Section \ref{futureresearch}).

\hypertarget{casestudy}{%
\subsection{Case Study}\label{casestudy}}
\subsubsection{Interpretation}
The explosion and subsequent sinking of the \emph{Deepwater Horizon}
drilling rig occurred in late April 2010, and by early June oil had
reached the shores of all five Gulf coast states. Photo-ID surveys
did not begin until 18 June, so we did not have information on the
population size or survival rates prior to the spill. For the first two years post-spill,
total population abundance steadily increased (Figure \ref{fig:salplot}). Over the same
time period, annual survival probability was initially low and then increased
(Figure \ref{fig:survplot}). When we compare annual survival estimates in 2010 in Barataria Bay (0.86) to the annual survival probabilities of similar
dolphin populations (\cite{stolen_model_2003} $0.902$, \cite{speakman2010mark} $0.951$, \cite{ludwig_survival_2021} $0.94$, \cite{wells_life_2025} $0.928-0.968$), it is evident that survival in Barataria Bay was low relative to other populations. Despite the the wide confidence interval on the first inter-primary interval survival estimate, over $90$\% of bootstrap resamples estimated survival to be increasing between the first and second inter-primary interval and between the second and third inter-primary interval (Appendix 8), providing strong evidence that survival was increasing over this period. Recruitment rate was
relatively high over the first two years post-spill (Figure \ref{fig:survplot}). Together,
these inferences are coherent with individuals entering the study area who
contribute to a higher survival rate compared to those present in the
study area immediately post-spill. These recruits could have been from outside
the study area or could have been previously non-distinct individuals present in the surveyed area that
became more distinctive. This is the first example of how interpretation must be
made in light of the model assumptions and how the data were collected.
Prior to \(2019\), the southeast stratum was not surveyed and so
individuals in this area prior to \(2019\) were likely to remain
undetected. This means that individuals that permanently immigrated from the southeast stratum to other strata prior to 2019 would be indistinguishable from individuals that permanently immigrated from anywhere outside the study area and the model would capture this
through the recruitment process. Therefore, when interpreting the results, the southeast
stratum should be considered outside
the study area prior to 2019. In other words, the high recruitment could reflect a
redistribution of individuals within the bay where individuals in
peripheral habitat replace those in island, central, and west strata who
were subject to high mortality.

Between \(2012\) and \(2014\), abundance was relatively stable
(Figure \ref{fig:salplot}). Annual survival probability decreased from \(2012\) to \(2013\)
and appears to have remained stable afterward with a mean \(\sim 0.9\) (Figure
\ref{fig:survplot}). This mortality did not affect abundance
substantially, likely because recruitment increased in $2013-2014$ (Figure \ref{fig:survplot}). The reason for this
increase in recruitment is not clear from the model: it could possibly
reflect further redistribution of individuals within the bay, increased
attribution of distinctive marks, or changes in prey distribution or fishing leading
to new individuals entering the population. Overall, the inference can
be interpreted in terms of change observed in the study: the population
continued to experience higher mortality than expected for comparable
dolphin populations and abundance did not reflect this substantially
due to a subsequent increase in new individuals entering the surveyed
area, an area that itself shifts in definition over primary occasions.  

Finally, during the \(5\) years between the \(2014\) survey and new
surveys in \(2019\), inference clearly points to a population where
recruitment was low and survival remained at its previous level, leading
to a decline in population abundance (Figure \ref{fig:salplot}).

Together with these temporal inferences, animal density surface estimation
allows us to infer where in the bay the population is concentrated and therefore where these events may have had the most impact. Based on the animal density surface, it is clear
that individuals spend much of their time around the islands and in
particular near the passes between the islands (Figure
\ref{fig:meanD_nocut}). This is likely due to the increased availability
of prey near the inlets of the Gulf \citep{hornsby2017using}. However, the highest animal densities were predicted outside the barrier island. The relative detectability of ACs outside the island is moderate, so this density may be attributable to the high number of detections in the pass at the northeast point of the island (see panel A of Supplementary Figure 9). The number of detections in the pass is much higher than in the area just inland of the barrier island, so the model is likely smoothing the associated ACs out into the area with less survey effort. Without survey effort beyond the barrier island, it is difficult to determine if the density estimate in this area is reliable or merely an artifact of our smoothing method.

\subsubsection{Assumptions}

OpenSCR models assume that individuals are detected independently with respect to their ACs. Dolphins are highly social, and simulations by \cite{lopez-bao_toward_2018} and \cite{bischof_consequences_2020} suggest that relatively high levels of social grouping behavior may cause problems with parameter estimation. Dolphin group sizes in the capture history spanned 1 to 58 distinctively marked individuals, with median group size of 3.  These are much smaller than the group sizes found to be problematic in the above studies.  For example, \cite{bischof_consequences_2020} found that even when group sizes are $6.25\%$ of the population and cohesion reached $50\%$, coverage (i.e., the probability that the $95\%$ confidence interval of the parameter estimate contains the true value) of density ($D$) and movement ($\sigma$) were $80\%$ and $85\%$ respectively. This would correspond to a group of 156 dolphins in our population, 22 times higher than our mean observed group size of 7 animals and 52 times higher than the median.  We therefore concluded that aggregation and cohesion of individuals is low enough that it does not significantly violate assumptions of detection independence between individuals. 

OpenSCR also assumes that all individuals have equal probabilities of detection and survival with respect to their ACs (and relevant covariates). Though sex information existed for a small subset of our identified individuals from biopsies or captures for health assessments, it is not usually possible to determine the sex of dolphins from field observations (aside from inferring that individuals accompanied by calves are likely female). We therefore did not include sex as a covariate for detection probabilities or population dynamics. While there is evidence that male survival is slightly lower than female survival \citep{schwacke2017quantifying}, we assume that our model estimates an average survival for the whole population. There is also some evidence that male dolphins range more widely than female dolphins \citep{wells2017ranging}. Because goodness-of-fit metrics did not reveal poor detection specification, we assume that estimating a population average movement parameter was sufficient. If this had not been the case, this could have likely been addressed with a two-part mixture model on detection parameters.

\hypertarget{limitations}{%
\subsection{Methods}\label{limitations}}

The strategies presented in this work for openSCR in a maximum likelihood framework (i.e., spatially-varying density, time-varying population dynamics, the use of search-encounter data, non-Euclidean distances, animal density surfaces, and multi-model inference) address the real-world challenges of generating inference from large, complex datasets. This case study demonstrated both advantages and challenges of the proposed openSCR method.

  \textbf{MLE computation time for multimodel inference and increased model complexity.} Our model presents the first implementation known to these authors of an open population SCR model with spatially-varying density and time-varying population dynamics in a maximum likelihood framework. While there are such SCR models in a Bayesian framework that are similarly large in scope (i.e., number of individuals, occasions, and traps; e.g., \cite{bischof_estimating_2020}), our use of MLE means there is only a small additional performance cost for increasing model complexity. For example, allowing ACs to move only required a recalculation of density for the marginalization rather than resampling of AC centers for the entire augmented superpopulation for each primary. The use of MLE allows for a much faster computation time for a single model (benchmarks provided in \cite{glennie2019open}; see also Appendix 9). This enabled us to compare multiple population dynamics models, density models, and detection parameter models by AIC, and to incorporate model-averaging to account for uncertainty in the hyper-parameter specifying the number of basis functions in the thin-plate regression splines. This in turn allowed for richer inference (including model-averaged smoothed estimates of population dynamics over time) than previous analyses by \cite{mcdonald2017survival} (Appendix 9), or than would have been possible with a comparable model in a Bayesian framework.
  
\textbf{Accounting for irregular primaries.} Because our model estimates continuous survival and recruitment rates from which probabilities and numbers of recruits are calculated for each inter-primary period, we were able to incorporate data from 2019 into our model. Estimates of survival and recruitment over the inter-primary period between April 2014 and March 2019 can be reliably interpreted as the average annual rates over this period. Note that this does not necessarily mean that, for example, annual survival probability was exactly 0.90 for all years from 2015-2019: it is possible that survival fluctuated around this mean value. Because we do not have capture history data during that time period, we are unable to infer fluctuations in parameters over the period. 

  \textbf{Sensitivity of inference to model selection.} Selecting
  between survival models with different hyper-parameters led to
  substantially different inference on how survival varied over time (Figure 3 shows that a model with 3 basis functions estimates little variation over the inter-primary intervals 1 through 6) and different initial survival estimates. Because we averaged over models with differing numbers of basis functions, the confidence interval on survival in the first inter-primary interval reflects the breadth of behaviors of the spline parameter surface.
  When assessing the impact of the oil spill, a justified estimate of
  survival in the first year post-spill was an important quantity \citep{schwacke2017quantifying}, and the sensitivity of this estimate
  to the smoothness of the underlying regression spline is therefore an
  important limitation to consider.
  We recommend three strategies for dealing with this sensitivity. (1) Inspect a model with survival rate as a function of primary as a factor. This can give some insight into the pattern underlying the smooth. For example, in Supplementary Figure 10, it is evident primary-specific estimates of survival vary widely: survival in early 2012 is estimated to be $100\%$. Note that AIC still prefers the smooth model, which is able to approximately describe this pattern with fewer parameters. (2) If multiple hyper-parameter variations of the best model produce models within 2 AIC units of the best performing model, use bootstrap samples to produce a model averaged estimate as we do here. Indeed, this model averaged estimate is very close to the estimate from the model with primary specific survival. This approach will, however, increase the uncertainty in areas where the smooth behavior is more variable; in our case, this is the first primary. (3) Given this additional variation, it can be useful to inspect the individual bootstrap estimates to elucidate consistent patterns. For example, Appendix 9 demonstrates that despite the wide confidence interval on the first inter-primary interval survival estimate, this estimate is always less than the survival estimates for the second and third inter-primary intervals, which indicates that while we may be uncertain about the exact survival probability in late 2011, we have high confidence of a pattern of increasing survival probabilities. 
  
  \textbf{Estimating smoothness.} Related to the previous limitation,
  using AIC to select the number of basis functions for regression splines may be a
  sub-optimal strategy in traditional smoothing spline applications. By
  limiting the number of basis functions, we likely over-smoothed density in
  areas where sharp changes exist, which is supported by the identified
  lack of fit in these areas.

\hypertarget{futureresearch}{%
\subsection{Research Opportunities}\label{futureresearch}}

Our proposed workflow focuses on what is currently feasible when
applying openSCR to large datasets in a maximum likelihood framework. Data collected over long time
periods and over large populations will, however, likely provide the
opportunity to infer more about populations using openSCR and motivate
the need for more realistic model components to be added. There are
clear opportunities for future research, as follows. 
    
\noindent \textbf{Estimating smoothness.} Rather than the stepwise AIC model selection used here, one could estimate smoothness of density surfaces, survival, and recruitment by shrinkage methods in a penalized likelihood framework \citep{wood2016smoothing}.

\noindent \textbf{Population Dynamics.} To increase biological realism, one could combine a population dynamics model that directly estimates vital rates with openSCR. While openSCR models provide estimates of apparent recruitment and survival rates based on patterns in the capture history, they do not explicitly model population dynamics. Inference therefore requires careful interpretation (e.g., estimated increases in recruitment could be from the movement of animals into the area or from individuals already present acquiring distinctive marks). Integrated population models  \citep[IPMs,][]{frost_integrated_2022} have allowed explicit estimation of population dynamics in capture-recapture studies \citep[such as age-dependent survival rates,][]{bird_combining_2018}. While IPMs have improved estimate precision in SCR \citep{chandler_spatially_2014}, they have not yet been used for explicit vital rate estimation, and could be used to examine the relative impact of pollution on the survival of different age classes.

\noindent \textbf{Heterogeneity.} One could incorporate further individual
heterogeneity by discrete or continuous random effects, e.g., to
capture those individuals affected more by the spill and so having
higher mortality, \citep{pledger2000unified}, or allow vital rates to vary through space. Individual heterogeneity in survival has been successfully implemented in a Bayesian framework \citep{honeycutt_spatial_2019, satter_sex-specific_2019} including spatially varying survival \citep{milleret_estimating_2023}, and these extensions could provide more specific inference about the effects of localized events on survival.

\noindent \textbf{Continuous observation process.} The model presented here discretized observation process onto a grid of traps. An alternative would be to construct a continuous-space, continuous-time detection process model that would better reflect the nature of the transect-based photo-ID survey and would avoid the need to decide on a trap grid spacing. While alternative methods for using transect data in open population SCR have been developed \citep{gowan_open_2021}, none have yet modelled a moving detector.

\hypertarget{data-availability}{%
\section*{Data Availability}\label{data-availability}}

The data and code to run the analyses will be available on Dryad.

\section*{Acknowledgments}

Not included while paper under review, except to say that we thank two anonymous reviewers for their thoughtful comments, which led to a greatly improved manuscript.

\bibliography{references.bib}

\begin{thebibliography}{}

\bibitem[Bischof et~al., 2020]{bischof_estimating_2020}
Bischof, R., Milleret, C., Dupont, P., Chipperfield, J., Tourani, M., Ordiz, A., de~Valpine, P., Turek, D., Royle, J.~A., Gimenez, O., Flagstad, O., \r{A}kesson, M., Svensson, L., Brøseth, H., and Kindberg, J. (2020).
\newblock Estimating and forecasting spatial population dynamics of apex predators using transnational genetic monitoring.
\newblock {\em Proceedings of the National Academy of Sciences}, 117(48):30531--30538.

\bibitem[Glennie et~al., 2019]{glennie2019open}
Glennie, R., Borchers, D.~L., Murchie, M., Harmsen, B.~J., and Foster, R.~J. (2019).
\newblock Open population maximum likelihood spatial capture-recapture.
\newblock {\em Biometrics}, 75(4):1345--1355.

\bibitem[Hornsby et~al., 2017]{hornsby_using_2017}
Hornsby, F., McDonald, T., Balmer, B., Speakman, T., Mullin, K., Rosel, P., Wells, R., Telander, A., Marcy, P., and Schwacke, L. (2017).
\newblock Using salinity to identify common bottlenose dolphin habitat in {Barataria} {Bay}, {Louisiana}, {USA}.
\newblock {\em Endangered Species Research}, 33:181--192.

\bibitem[Royle et~al., 2007]{royle2007_analysisofmulti}
Royle, J., Dorazio, R., and Link, W. (2007).
\newblock Analysis of multinomial models with unknown index using data augmentation.
\newblock {\em Journal of Computational and Graphical Statistics}, 16:67–85.

\bibitem[Takeshita et~al., 2021]{takeshita2021high}
Takeshita, R., Balmer, B.~C., Messina, F., Zolman, E.~S., Thomas, L., Wells, R.~S., Smith, C.~R., Rowles, T.~K., and Schwacke, L.~H. (2021).
\newblock High site-fidelity in common bottlenose dolphins despite low salinity exposure and associated indicators of compromised health.
\newblock {\em PLOS ONE}, 16(9):e0258031.

\bibitem[Wells et~al., 2017]{wells_ranging_2017}
Wells, R., Schwacke, L., Rowles, T., Balmer, B., Zolman, E., Speakman, T., Townsend, F., Tumlin, M., Barleycorn, A., and Wilkinson, K. (2017).
\newblock Ranging patterns of common bottlenose dolphins {Tursiops} truncatus in {Barataria} {Bay}, {Louisiana}, following the {Deepwater} {Horizon} oil spill.
\newblock {\em Endangered Species Research}, 33:159--180.

\bibitem[White et~al., 2018]{white2018salinity}
White, E.~D., Messina, F., Moss, L., and Meselhe, E. (2018).
\newblock Salinity and marine mammal dynamics in barataria basin: Historic patterns and modeled diversion scenarios.
\newblock {\em Water}, 10(8):1015.

\bibitem[Wood et~al., 2016]{wood2016smoothing}
Wood, S.~N., Pya, N., and S{\"a}fken, B. (2016).
\newblock Smoothing parameter and model selection for general smooth models.
\newblock {\em Journal of the American Statistical Association}, 111(516):1548--1563.

\end{thebibliography}


\begin{thebibliography}{}

\bibitem[Akaike, 1998]{akaike1998information}
Akaike, H. (1998).
\newblock Information theory and an extension of the maximum likelihood principle.
\newblock In {\em Selected Papers of {H}irotugu {A}kaike}, pages 199--213. Springer.

\bibitem[Augustine et~al., 2020]{augustine_sex-specific_2020}
Augustine, B.~C., Kéry, M., Olano~Marin, J., Mollet, P., Pasinelli, G., and Sutherland, C. (2020).
\newblock Sex-specific population dynamics and demography of capercaillie ({Tetrao} urogallus {L}.) in a patchy environment.
\newblock {\em Population Ecology}, 62(1):80--90.

\bibitem[Bird et~al., 2018]{bird_combining_2018}
Bird, T., Lyon, J., Wotherspoon, S., Todd, C., Tonkin, Z., and McCarthy, M. (2018).
\newblock Combining capture–recapture data and known ages allows estimation of age‐dependent survival rates.
\newblock {\em Ecology and Evolution}, 9(1):90--99.

\bibitem[Bischof et~al., 2016]{bischof_wildlife_2016}
Bischof, R., Brøseth, H., and Gimenez, O. (2016).
\newblock Wildlife in a {Politically} {Divided} {World}: {Insularism} {Inflates} {Estimates} of {Brown} {Bear} {Abundance}.
\newblock {\em Conservation Letters}, 9(2):122--130.

\bibitem[Bischof et~al., 2020a]{bischof_consequences_2020}
Bischof, R., Dupont, P., Milleret, C., Chipperfield, J., and Royle, J.~A. (2020a).
\newblock Consequences of ignoring group association in spatial capture–recapture analysis.
\newblock {\em Wildlife Biology}, 2020(1):wlb.00649.

\bibitem[Bischof et~al., 2020b]{bischof_estimating_2020}
Bischof, R., Milleret, C., Dupont, P., Chipperfield, J., Tourani, M., Ordiz, A., de~Valpine, P., Turek, D., Royle, J.~A., Gimenez, O., Flagstad, O., \r{A}kesson, M., Svensson, L., Brøseth, H., and Kindberg, J. (2020b).
\newblock Estimating and forecasting spatial population dynamics of apex predators using transnational genetic monitoring.
\newblock {\em Proceedings of the National Academy of Sciences}, 117(48):30531--30538.

\bibitem[Borchers and Efford, 2008]{borchers2008spatially}
Borchers, D.~L. and Efford, M.~G. (2008).
\newblock Spatially explicit maximum likelihood methods for capture--recapture studies.
\newblock {\em Biometrics}, 64(2):377--385.

\bibitem[Chandler and Clark, 2014]{chandler_spatially_2014}
Chandler, R.~B. and Clark, J.~D. (2014).
\newblock Spatially explicit integrated population models.
\newblock {\em Methods in Ecology and Evolution}, 5(12):1351--1360.
\newblock \_eprint: https://onlinelibrary.wiley.com/doi/pdf/10.1111/2041-210X.12153.

\bibitem[Chandler et~al., 2018]{chandler_characterizing_2018}
Chandler, R.~B., Hepinstall-Cymerman, J., Merker, S., Abernathy-Conners, H., and Cooper, R.~J. (2018).
\newblock Characterizing spatio-temporal variation in survival and recruitment with integrated population models.
\newblock {\em The Auk}, 135(3):409--426.

\bibitem[Cormack, 1964]{cormack1964estimates}
Cormack, R.~M. (1964).
\newblock Estimates of survival from the sighting of marked animals.
\newblock {\em Biometrika}, 51(3/4):429--438.

\bibitem[Diggle, 2013]{diggle2013statistical}
Diggle, P.~J. (2013).
\newblock {\em Statistical analysis of spatial and spatio-temporal point patterns}.
\newblock CRC press.

\bibitem[Durbach et~al., 2024]{durbach_thats_2024}
Durbach, I., Chopara, R., Borchers, D.~L., Phillip, R., Sharma, K., and Stevenson, B.~C. (2024).
\newblock That's not the {Mona} {Lisa}! {How} to interpret spatial capture-recapture density surface estimates.
\newblock {\em Biometrics}, 80(1):ujad020.

\bibitem[Efford and Schofield, 2020]{efford2020spatial}
Efford, M.~G. and Schofield, M.~R. (2020).
\newblock A spatial open-population capture-recapture model.
\newblock {\em Biometrics}, 76(2):392--402.

\bibitem[Ergon and Gardner, 2014a]{ergon2014separating}
Ergon, T. and Gardner, B. (2014a).
\newblock Separating mortality and emigration: modelling space use, dispersal and survival with robust-design spatial capture--recapture data.
\newblock {\em Methods in Ecology and Evolution}, 5(12):1327--1336.

\bibitem[Ergon and Gardner, 2014b]{ergon_separating_2014}
Ergon, T. and Gardner, B. (2014b).
\newblock Separating mortality and emigration: modelling space use, dispersal and survival with robust-design spatial capture–recapture data.
\newblock {\em Methods in Ecology and Evolution}, 5(12):1327--1336.

\bibitem[Friday et~al., 2000]{friday_measurement_2000}
Friday, N., Smith, T.~D., Stevick, P.~T., and Allen, J. (2000).
\newblock Measurement of {Photographic} {Quality} and {Individual} {Distinctiveness} for the {Photographic} {Identification} of {Humpback} {Whales}, {Megaptera} {Novaeangliae}.
\newblock {\em Marine Mammal Science}, 16(2):355--374.

\bibitem[Frost et~al., 2022]{frost_integrated_2022}
Frost, F., McCrea, R., King, R., Gimenez, O., and Zipkin, E. (2022).
\newblock Integrated {Population} {Models}: {Achieving} {Their} {Potential}.
\newblock {\em Journal of Statistical Theory and Practice}, 17(1):6.

\bibitem[Gardner et~al., 2010]{gardner2010spatially}
Gardner, B., Reppucci, J., Lucherini, M., and Royle, J.~A. (2010).
\newblock Spatially explicit inference for open populations: estimating demographic parameters from camera-trap studies.
\newblock {\em Ecology}, 91(11):3376--3383.

\bibitem[Gardner et~al., 2018]{gardner2018state}
Gardner, B., Sollmann, R., Kumar, N.~S., Jathanna, D., and Karanth, K.~U. (2018).
\newblock State space and movement specification in open population spatial capture--recapture models.
\newblock {\em Ecology and Evolution}, 8(20):10336--10344.

\bibitem[Garrison et~al., 2020]{garrison2020predicting}
Garrison, L.~P., Litz, J., and Sinclair, C. (2020).
\newblock Predicting the effects of low salinity associated with the {MBSD} project on resident common bottlenose dolphins ({Tursiops} truncatus) in {Barataria} {Bay}, {LA}.
\newblock {NOAA} {Technical} {Memorandum} NMFS-SEFSC, National Oceanic and Atmospheric Administration ({NOAA}).

\bibitem[Glennie et~al., 2019]{glennie2019open}
Glennie, R., Borchers, D.~L., Murchie, M., Harmsen, B.~J., and Foster, R.~J. (2019).
\newblock Open population maximum likelihood spatial capture-recapture.
\newblock {\em Biometrics}, 75(4):1345--1355.

\bibitem[Gowan et~al., 2021]{gowan_open_2021}
Gowan, T.~A., Crum, N.~J., and Roberts, J.~J. (2021).
\newblock An open spatial capture–recapture model for estimating density, movement, and population dynamics from line-transect surveys.
\newblock {\em Ecology and Evolution}, 11(12):7354--7365.

\bibitem[Hayes et~al., 2018]{hayes2018us}
Hayes, S.~A., Josephson, E., Maze-Foley, K., and Rosel, P.~E. (2018).
\newblock {US} {A}tlantic and {G}ulf of {M}exico marine mammal stock assessments-2017.
\newblock Technical report, NOAA.

\bibitem[Honeycutt et~al., 2019]{honeycutt_spatial_2019}
Honeycutt, R.~K., Garwood, J.~M., Lowe, W.~H., and Hossack, B.~R. (2019).
\newblock Spatial capture–recapture reveals age- and sex-specific survival and movement in stream amphibians.
\newblock {\em Oecologia}, 190(4):821--833.

\bibitem[Hornsby et~al., 2017]{hornsby2017using}
Hornsby, F.~E., McDonald, T.~L., Balmer, B.~C., Speakman, T.~R., Mullin, K.~D., Rosel, P.~E., Wells, R.~S., Telander, A.~C., Marcy, P.~W., Klaphake, K.~C., and Schwacke, L.~H. (2017).
\newblock Using salinity to identify common bottlenose dolphin habitat in {B}arataria {B}ay, {L}ouisiana, {USA}.
\newblock {\em Endangered Species Research}, 33:181--192.

\bibitem[Jolly, 1965]{jolly1965explicit}
Jolly, G.~M. (1965).
\newblock Explicit estimates from capture-recapture data with both death and immigration-stochastic model.
\newblock {\em Biometrika}, 52(1/2):225--247.

\bibitem[Kendall et~al., 1997]{kendall1997estimating}
Kendall, W.~L., Nichols, J.~D., and Hines, J.~E. (1997).
\newblock Estimating temporary emigration using capture--recapture data with {P}ollock’s robust design.
\newblock {\em Ecology}, 78(2):563--578.

\bibitem[Ludwig et~al., 2021]{ludwig_survival_2021}
Ludwig, K.~E., Daly, M., Levesque, S., and Berrow, S.~D. (2021).
\newblock Survival {Rates} and {Capture} {Heterogeneity} of {Bottlenose} {Dolphins} ({Tursiops} truncatus) in the {Shannon} {Estuary}, {Ireland}.
\newblock {\em Frontiers in Marine Science}, 8.

\bibitem[López-Bao et~al., 2018]{lopez-bao_toward_2018}
López-Bao, J.~V., Godinho, R., Pacheco, C., Lema, F.~J., García, E., Llaneza, L., Palacios, V., and Jiménez, J. (2018).
\newblock Toward reliable population estimates of wolves by combining spatial capture-recapture models and non-invasive {DNA} monitoring.
\newblock {\em Scientific Reports}, 8(1):2177.

\bibitem[Manly, 2006]{manly2006randomization}
Manly, B.~F. (2006).
\newblock {\em Randomization, bootstrap and Monte Carlo methods in biology}.
\newblock CRC Press.

\bibitem[McDonald et~al., 2017]{mcdonald2017survival}
McDonald, T.~L., Hornsby, F.~E., Speakman, T.~R., Zolman, E.~S., Mullin, K.~D., Sinclair, C., Rosel, P.~E., Thomas, L., and Schwacke, L.~H. (2017).
\newblock Survival, density, and abundance of common bottlenose dolphins in {B}arataria {B}ay ({USA}) following the {D}eepwater {H}orizon oil spill.
\newblock {\em Endangered Species Research}, 33:193--209.

\bibitem[Melancon et~al., 2011]{melancon2011photo}
Melancon, R.~A., Lane, S., Speakman, T., Hart, L.~B., Sinclair, C., Adams, J., Rosel, P.~E., and Schwacke, L. (2011).
\newblock Photo-identification field and laboratory protocols utilizing finbase version 2.
\newblock Technical report, NOAA.

\bibitem[Milleret et~al., 2023]{milleret_estimating_2023}
Milleret, C., Dey, S., Dupont, P., Brøseth, H., Turek, D., de~Valpine, P., and Bischof, R. (2023).
\newblock Estimating spatially variable and density-dependent survival using open-population spatial capture–recapture models.
\newblock {\em Ecology}, 104(2):e3934.

\bibitem[Pledger, 2000]{pledger2000unified}
Pledger, S. (2000).
\newblock Unified maximum likelihood estimates for closed capture--recapture models using mixtures.
\newblock {\em Biometrics}, 56(2):434--442.

\bibitem[Pollock, 1982]{pollock1982capture}
Pollock, K.~H. (1982).
\newblock A capture-recapture design robust to unequal probability of capture.
\newblock {\em The Journal of Wildlife Management}, 46(3):752--757.

\bibitem[Royle et~al., 2013a]{royle_spatial_2013}
Royle, J.~A., Chandler, R.~B., Gazenski, K.~D., and Graves, T.~A. (2013a).
\newblock Spatial capture–recapture models for jointly estimating population density and landscape connectivity.
\newblock {\em Ecology}, 94(2):287--294.

\bibitem[Royle et~al., 2013b]{royle2013spatial}
Royle, J.~A., Chandler, R.~B., Sollmann, R., and Gardner, B. (2013b).
\newblock {\em Spatial capture-recapture}.
\newblock Academic Press.

\bibitem[Royle et~al., 2013c]{royle_integrating_2013}
Royle, J.~A., Chandler, R.~B., Sun, C.~C., and Fuller, A.~K. (2013c).
\newblock Integrating resource selection information with spatial capture–recapture.
\newblock {\em Methods in Ecology and Evolution}, 4(6):520--530.

\bibitem[Royle et~al., 2011]{royle2011spatial}
Royle, J.~A., Kery, M., and Guelat, J. (2011).
\newblock Spatial capture-recapture models for search-encounter data.
\newblock {\em Methods in Ecology and Evolution}, 2(6):602--611.

\bibitem[Satter et~al., 2019]{satter_sex-specific_2019}
Satter, C.~B., Augustine, B.~C., Harmsen, B.~J., Foster, R.~J., and Kelly, M.~J. (2019).
\newblock Sex-specific population dynamics of ocelots in {Belize} using open population spatial capture–recapture.
\newblock {\em Ecosphere}, 10(7):e02792.

\bibitem[Schwacke et~al., 2017]{schwacke2017quantifying}
Schwacke, L.~H., Thomas, L., Wells, R.~S., McFee, W.~E., Hohn, A.~A., Mullin, K.~D., Zolman, E.~S., Quigley, B.~M., Rowles, T.~K., and Schwacke, J.~H. (2017).
\newblock Quantifying injury to common bottlenose dolphins from the {D}eepwater {H}orizon oil spill using an age-, sex-and class-structured population model.
\newblock {\em Endangered Species Research}, 33:265--279.

\bibitem[Schwarz and Arnason, 1996]{schwarz1996general}
Schwarz, C.~J. and Arnason, A.~N. (1996).
\newblock A general methodology for the analysis of capture-recapture experiments in open populations.
\newblock {\em Biometrics}, 52(3):860--873.

\bibitem[Seber and Schofield, 2019]{seber2019capture}
Seber, G. A.~F. and Schofield, M.~R. (2019).
\newblock {\em Capture-recapture: Parameter estimation for open animal populations}.
\newblock Springer.

\bibitem[Speakman et~al., 2010]{speakman2010mark}
Speakman, T.~R., Lane, S.~M., Schwacke, L.~H., Fair, P.~A., and Zolman, E.~S. (2010).
\newblock Mark-recapture estimates of seasonal abundance and survivorship for bottlenose dolphins (\textit{{T}ursiops truncatus}) near {C}harleston, {S}outh {C}arolina, {USA}.
\newblock {\em Journal of Cetacean Research and Management}, 11(2):153--162.

\bibitem[Stolen and Barlow, 2003]{stolen_model_2003}
Stolen, M.~K. and Barlow, J. (2003).
\newblock A {Model} {Life} {Table} for {Bottlenose} {Dolphins} (tursiops {Truncatus}) from the {Indian} {River} {Lagoon} {System}, {Florida}, {U}.s.a.
\newblock {\em Marine Mammal Science}, 19(4):630--649.

\bibitem[Takeshita et~al., 2017]{takeshita2017deepwater}
Takeshita, R., Sullivan, L., Smith, C., Collier, T., Hall, A., Brosnan, T., Rowles, T., and Schwacke, L. (2017).
\newblock The {D}eepwater {H}orizon oil spill marine mammal injury assessment.
\newblock {\em Endangered Species Research}, 33:95--106.

\bibitem[Urian et~al., 2015]{urian_recommendations_2015}
Urian, K., Gorgone, A., Read, A., Balmer, B., Wells, R.~S., Berggren, P., Durban, J., Eguchi, T., Rayment, W., and Hammond, P.~S. (2015).
\newblock Recommendations for photo-identification methods used in capture-recapture models with cetaceans.
\newblock {\em Marine Mammal Science}, 31(1):298--321.

\bibitem[Urian et~al., 1999]{urian_status_1999}
Urian, K.~W., Hohn, A.~A., and Hansen, L.~J. (1999).
\newblock Status of the photo-identification catalog of coastal bottlenose dolphins of the western north {Atlantic}: {Report} of a workshop of catalog contributors.
\newblock {NOAA} {Administrative} {Report} NMFS-SEFSC-425, National Oceanic and Atmospheric Administration ({NOAA}).

\bibitem[{van Etten}, 2017]{gdistance}
{van Etten}, J. (2017).
\newblock R package gdistance: Distances and routes on geographical grids.
\newblock {\em Journal of Statistical Software}, 76(13):1--21.

\bibitem[Wells et~al., 2025]{wells_life_2025}
Wells, R.~S., Hohn, A.~A., Scott, M.~D., Sweeney, J.~C., Townsend, F.~I., Allen, J.~B., Barleycorn, A.~A., McHugh, K.~A., Bassos-Hull, K., Lovewell, G.~N., Duffield, D.~A., Smith, C.~R., and Irvine, A.~B. (2025).
\newblock Life history, reproductive, and demographic parameters for bottlenose dolphins ({Tursiops} truncatus) in {Sarasota} {Bay}, {Florida}.
\newblock {\em Frontiers in Marine Science}, 12.

\bibitem[Wells et~al., 2017]{wells2017ranging}
Wells, R.~S., Schwacke, L.~H., Rowles, T.~K., Balmer, B.~C., Zolman, E., Speakman, T., Townsend, F.~I., Tumlin, M.~C., Barleycorn, A., and Wilkinson, K.~A. (2017).
\newblock Ranging patterns of common bottlenose dolphins \textit{{T}ursiops truncatus} in {B}arataria {B}ay, {L}ouisiana, following the {D}eepwater {H}orizon oil spill.
\newblock {\em Endangered Species Research}, 33:159--180.

\bibitem[Williams et~al., 2002]{williams2002analysis}
Williams, B.~K., Nichols, J.~D., and Conroy, M.~J. (2002).
\newblock {\em Analysis and management of animal populations}.
\newblock Academic Press.

\bibitem[Winton et~al., 2023]{winton_open_2023}
Winton, M.~V., Fay, G., and Skomal, G.~B. (2023).
\newblock An open spatial capture-recapture framework for estimating the abundance and seasonal dynamics of white sharks at aggregation sites.
\newblock {\em Marine Ecology Progress Series}, 715:1--25.

\bibitem[Wood, 2003]{wood2003thin}
Wood, S.~N. (2003).
\newblock Thin plate regression splines.
\newblock {\em Journal of the Royal Statistical Society: Series B (Statistical Methodology)}, 65(1):95--114.

\bibitem[Wood et~al., 2016]{wood2016smoothing}
Wood, S.~N., Pya, N., and S{\"a}fken, B. (2016).
\newblock Smoothing parameter and model selection for general smooth models.
\newblock {\em Journal of the American Statistical Association}, 111(516):1548--1563.

\end{thebibliography}

\newpage

\appendix{}

\title{Appendices to ``Estimating spatially-varying density and time-varying
demographics with open population spatial capture-recapture: a photo-ID
case study on bottlenose dolphins in Barataria Bay, Louisiana, USA''}
\author{
  }

\providecommand{\tightlist}{%
  \setlength{\itemsep}{0pt}\setlength{\parskip}{0pt}}

\renewcommand{\thesection}{Appendix \arabic{section}:}
\renewcommand{\figurename}{Supplementary Figure}
\renewcommand{\tablename}{Supplementary Table}

\maketitle

These appendices include supplementary figures, tables, and technical
details that are referred to in the main paper.

\hypertarget{appendix-3-data-summary}{%
\section{Capture History Summary}\label{appendix-3-capthist-summary}}

\begin{figure}[ht!]
    \includegraphics[width = 0.8\textwidth]{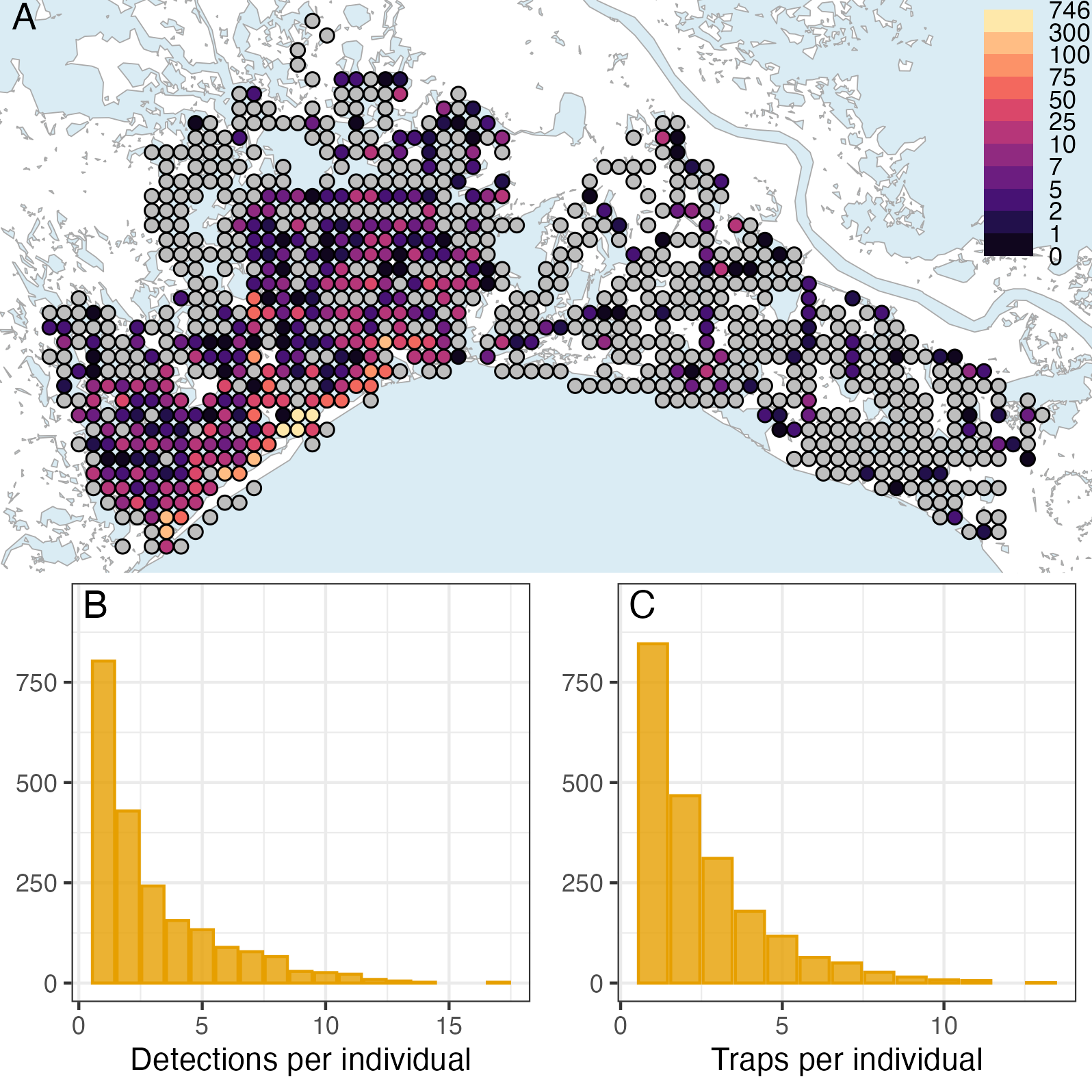}
    \caption{Individuals detected per trap (grey traps have no detections; A), frequencies of detections per individual (B), and frequencies of number of traps that detected an individual (C).}
    \label{fig:detections}
\end{figure}

General summaries of the capture history are presented here. The number of individuals detected per trap ranged from 0 to 746 over the course of the entire study (Supplementary Figure \ref{fig:detections} panel A). While 803 of the 2091 individuals were detected once, one individual was detected 19 times over 16 traps.  A graphical summary of the distribution of detections over space and the frequency of detections per individual and traps per individual is given in Supplementary Figure~\ref{fig:detections}.

\hypertarget{appendix-proportionmarked}{%
\section{Proportion Distinctive} \label{appendix-propmarked}}

For each sighting event, the total number of dolphins and the number of distinctive individuals were recorded. A total of 833 sighting events were recorded over the 34 secondary survey occasions.  We fit a suite of generalized additive models \citepapp[GAM]{wood2016smoothing} to these data to determine if the proportion of distinctive individuals changed over time (date of survey) or space ($x$, $y$ location). The response variable, number of distinct dolphins per sighting event, was assumed to follow a binomial distribution with number of trials equal to the total number of dolphins for that event. Probability of success (i.e., being distinctive) was linked to the linear predictor via a logit link.  Model specifications for the linear predictor and AIC score differences for all models considered are shown in Table \ref{tab:propdistinctmodels}. All models were fit with restricted maximum likelihood. The model with lowest AIC was a smooth thin-plate regression spline with shrinkage over $x$ and $y$. The predicted proportion distinctive surface is given in Figure 4 of the main text.

\begin{table}[ht!]
\centering
\begin{tabular}{p{7cm}p{5cm}l}
  \hline
 Model specification & Description & $\Delta$AIC \\ 
  \hline
 s(x, y, bs = "tp") & Smooth over $x$ and $y$ & 0.000 \\ 
  s(x, y, bs= ``tp'') + s(Date, bs = ``tp'') + ti(x, y, Date, d = c(2,1), bs = c(``tp'',``tp'')) & Main effect smooth over $x$ and $y$, main effect smooth over date, and tensor product interactive effect smooth over both & 0.899 \\ 
   s(Date, bs = "tp") + s(x, y, bs = "tp") & Smooth over $x$ and $y$ and smooth over date & 1.138 \\
   s(Date, bs = "tp") & Smooth over date & 123.463 \\
   1 & Constant & 124.984 \\
   \hline
\end{tabular}
\caption{Generalized additive model specification and $\Delta$AIC (difference between AIC for that model and the lowest AIC) for models of proportion of distinctive individuals per sighting event. Covariate ``Date'' refers to survey date, `x` and `y` to spatial location (in UTM); basis function ``tp'' means thin-plate regression spline.} 
\label{tab:propdistinctmodels}
\end{table}

\begin{figure}[tph]
    \includegraphics[width = 0.7\textwidth]{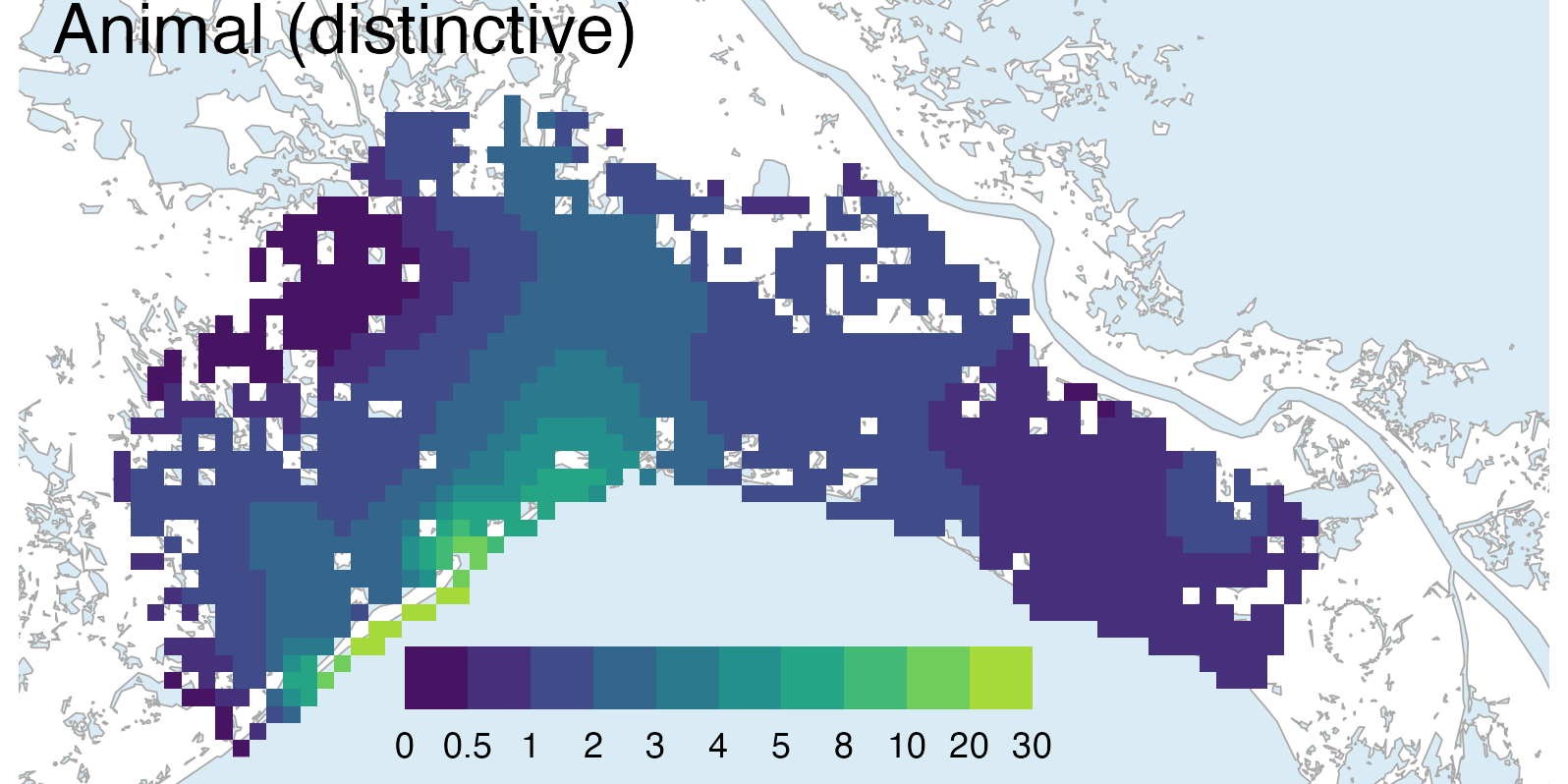}
    \caption{Derived mean alive, distinctive animal density ($D^{\alpha}$, animals per km$^2$) from $10000$ model-averaging bootstrap }
    \label{app-fig-animdensdist}
\end{figure}

\hypertarget{appendix-2-occasions}{%
\section{ Occasions}\label{appendix-2-occasions}}

The photo-ID surveys are first grouped into 34 secondary occasions and
then secondary occasions are grouped into 11 primary occasions. Each
primary occasion contained \(3\) secondary occasions, except the final
primary which contained \(4\). On average, each secondary occasion
contains \(5\) tracks. Supplementary Table \ref{tab:primstart} gives the start date of
each primary occasion.

\begin{table}[ht]
\centering
\begin{tabular}{rl}
  \hline
Primary & Start Date \\ 
  \hline
  1 & 18-Jun-10 \\ 
    2 & 09-Nov-10 \\ 
    3 & 06-Apr-11 \\ 
    4 & 09-Jun-11 \\ 
    5 & 09-Nov-11 \\ 
    6 & 07-Feb-12 \\ 
    7 & 11-Apr-12 \\ 
    8 & 09-Apr-13 \\ 
    9 & 09-Nov-13 \\ 
   10 & 22-Apr-14 \\ 
   11 & 14-Mar-19 \\ 
   \hline
\end{tabular}
\caption{Start date of each primary occasion} 
\label{tab:primstart}
\end{table}

The time between primary occasions (which is important for scaling the
relevant population dynamics, i.e., survival over one year is different
from survival over two months) was computed as the time between the
mid-points of each primary occasion given in units of years in Supplementary Table~\ref{tab:primdt}.

\begin{table}[ht]
\centering
\begin{tabular}{rrrr}
  \hline
Inter-primary & From primary & To primary & Duration \\ 
interval & occasion & occasion & (years) \\
  \hline
1 & 1 & 2 & 0.4 \\ 
2 & 2 & 3 & 0.4 \\ 
3 &  3 & 4 & 0.2 \\ 
4 &  4 & 5 & 0.4 \\ 
5 &  5 & 6 & 0.2 \\ 
6 &  6 & 7 & 0.2 \\ 
7 &  7 & 8 & 1.0 \\ 
8 &  8 & 9 & 0.6 \\ 
9 &  9 & 10 & 0.4 \\ 
10 &  10 & 11 & 4.9 \\ 
   \hline
\end{tabular}
\caption{Duration of time interval (in years and rounded to one decimal place) between primary occasions.} 
\label{tab:primdt}
\end{table}

\hypertarget{appendix-2-meshsim}{%
\section{Mesh Selection}\label{appendix-2-meshsim}}

Typical SCR analyses require a mesh buffer of $(3-4)\sigma$ from the traps. However, when a boundary restricts the area in which ACs can occur, the mesh should not be extended beyond that boundary (e.g., an island fully covered by a trap array). To demonstrate this, we sampled a population with constant AC density fully enclosed by a boundary, simulated 1000 capture histories from traps fully covering the enclosed area, then fit spatial capture recapture models with mesh buffer from $(0 - 4)\sigma$. As shown in Supplementary Figure \ref{app-fig-meshsim}, extending the mesh beyond the buffer results in increasingly negative bias in the AC density estimate. To avoid bias, the mesh should conform to the area in which ACs can occur. 
\begin{figure}[tph]
    \includegraphics[width = 0.7\textwidth]{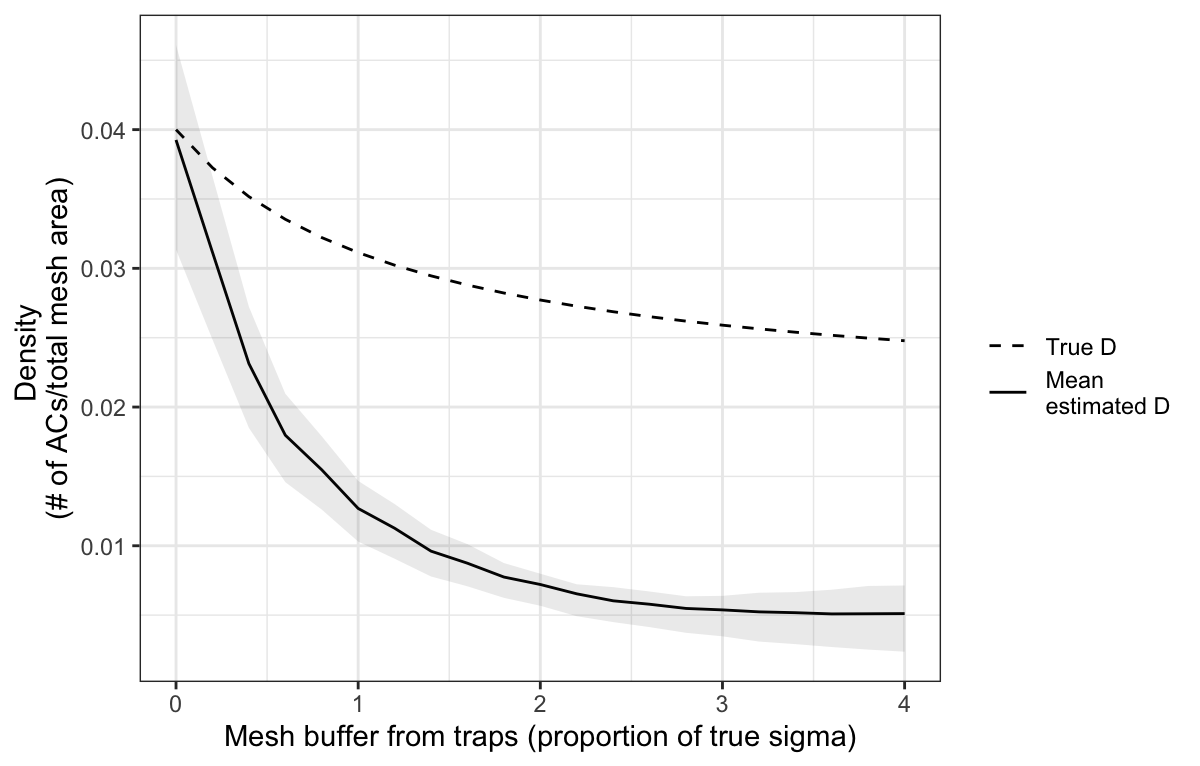}
    \caption{Results of 1000 simulated capture histories from populations with a constant AC density and subsequent SCR model AC density (D) estimates for a population contained within a boundary (solid line) and the $95\%$ confidence interval (grey ribbon). The dashed line is the true density of simulated ACs per total mesh area.}
    \label{app-fig-meshsim}
\end{figure}

A previous study fitted 44 dolphins in Barataria Bay with satellite-linked transmitters (\citepapp{wells_ranging_2017}) and found that none of the high quality locations during the study were recorded outside the area. The furthest recorded locations from the bay were within 4.24~km of shore, and $85.8\%$ of locations were inshore of the barrier islands \citepapp{wells_ranging_2017}. Utilization distributions from \cite{wells_ranging_2017} are primarily, if not exclusively, within the barrier islands. This is strong evidence that dolphins in Barataria Bay have ACs within the bay. 

Given that no dolphin ACs will occur beyond the barrier islands, we limited our mesh to $1$km beyond the barrier islands to avoid bias in the AC density estimate.

\hypertarget{appendix-3-covariates}{%
\section{Covariates}
\label{appendix-3-covariates}}

The analysis included the following possible covariates: time, stratum,
openness, average salinity, and spatial location \((x,y)\). Time refers to the
time at which each primary occasion took place measured in noninteger years. The
other covariates are spatially-referenced. Stratum and openness
are intended to explain variation in detectability across traps, while average salinity and  \((x,y)\) are intended to explain variation in density of activity centers. More information about the stratum, openness and salinity covariates are given below.

\begin{figure}[h]
\includegraphics[width = 0.8\textwidth]{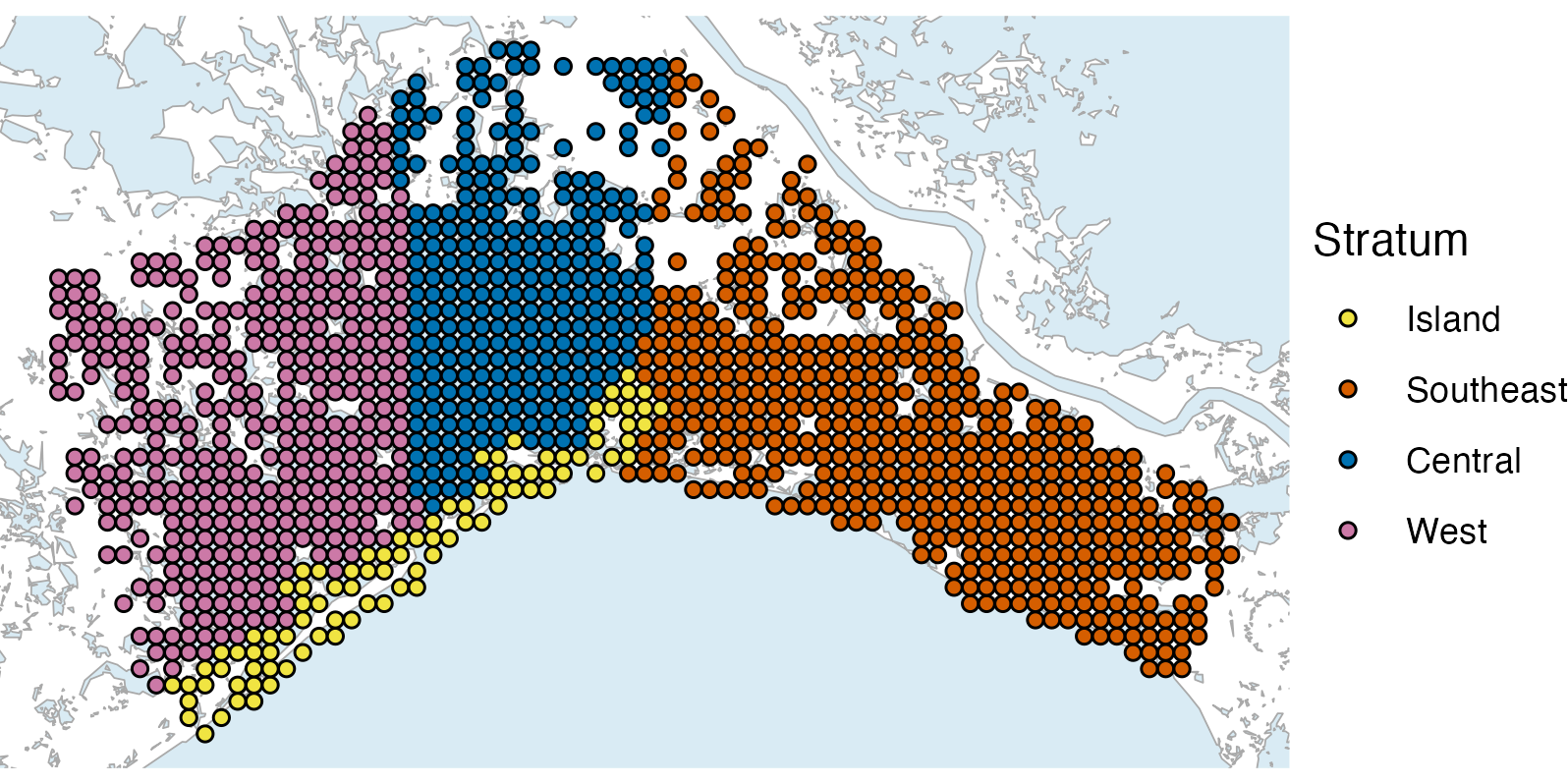} \caption{Strata}\label{fig:appstrata}
\end{figure}

Stratum was defined as a categorical variable that separated spatial regions that may have had different sighting rates (Supplementary Figure~\ref{fig:appstrata}).

\begin{figure}[h]
\includegraphics[width = 0.8\textwidth]{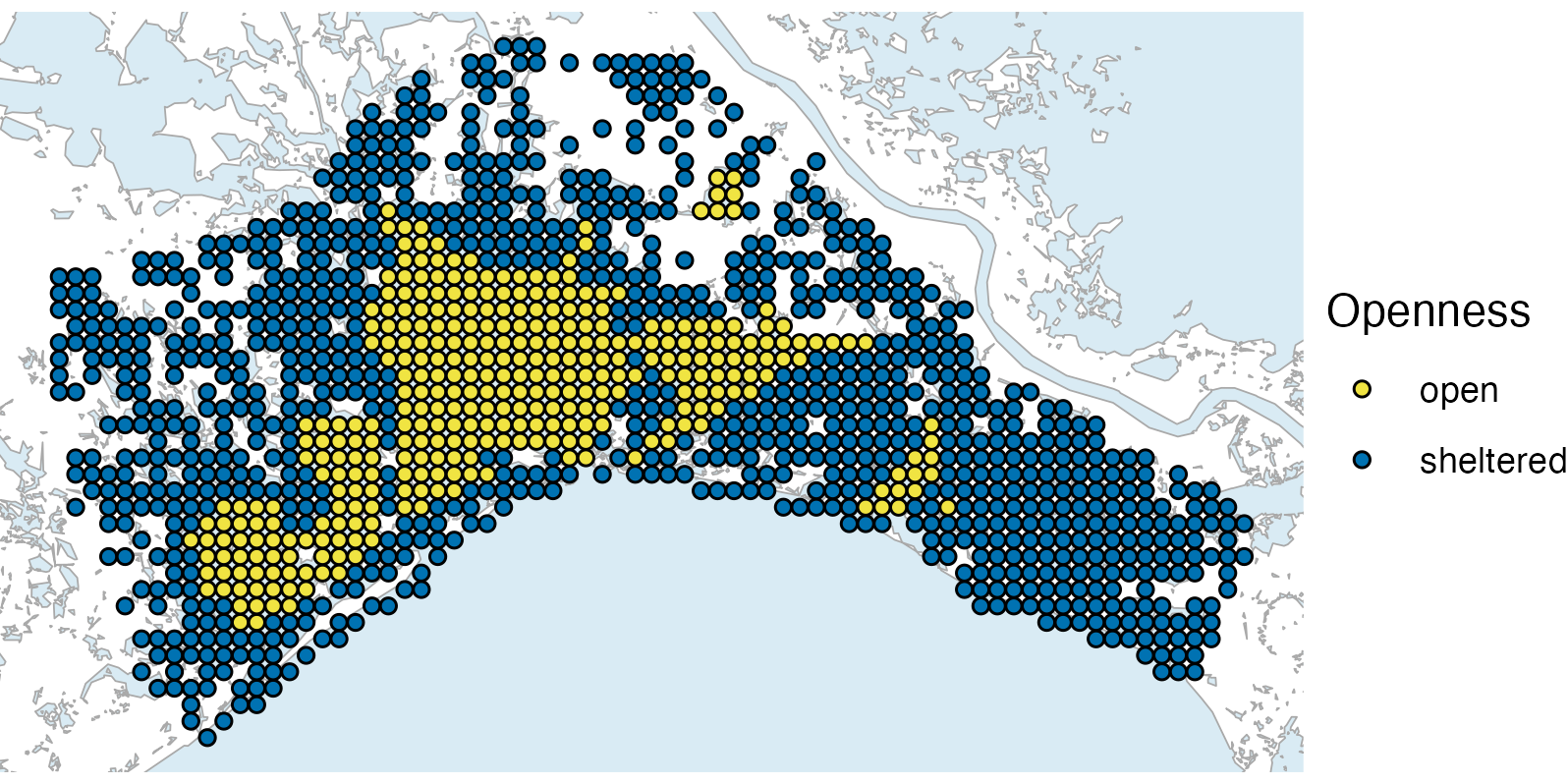} \caption{Openness}\label{fig:appopencov}
\end{figure}

Openness is another categorical variable used to capture the relatively
open and sheltered parts of the bay (Supplementary Figure~\ref{fig:appopencov}).\\

\begin{figure}[h]
\includegraphics[width = 0.8\textwidth]{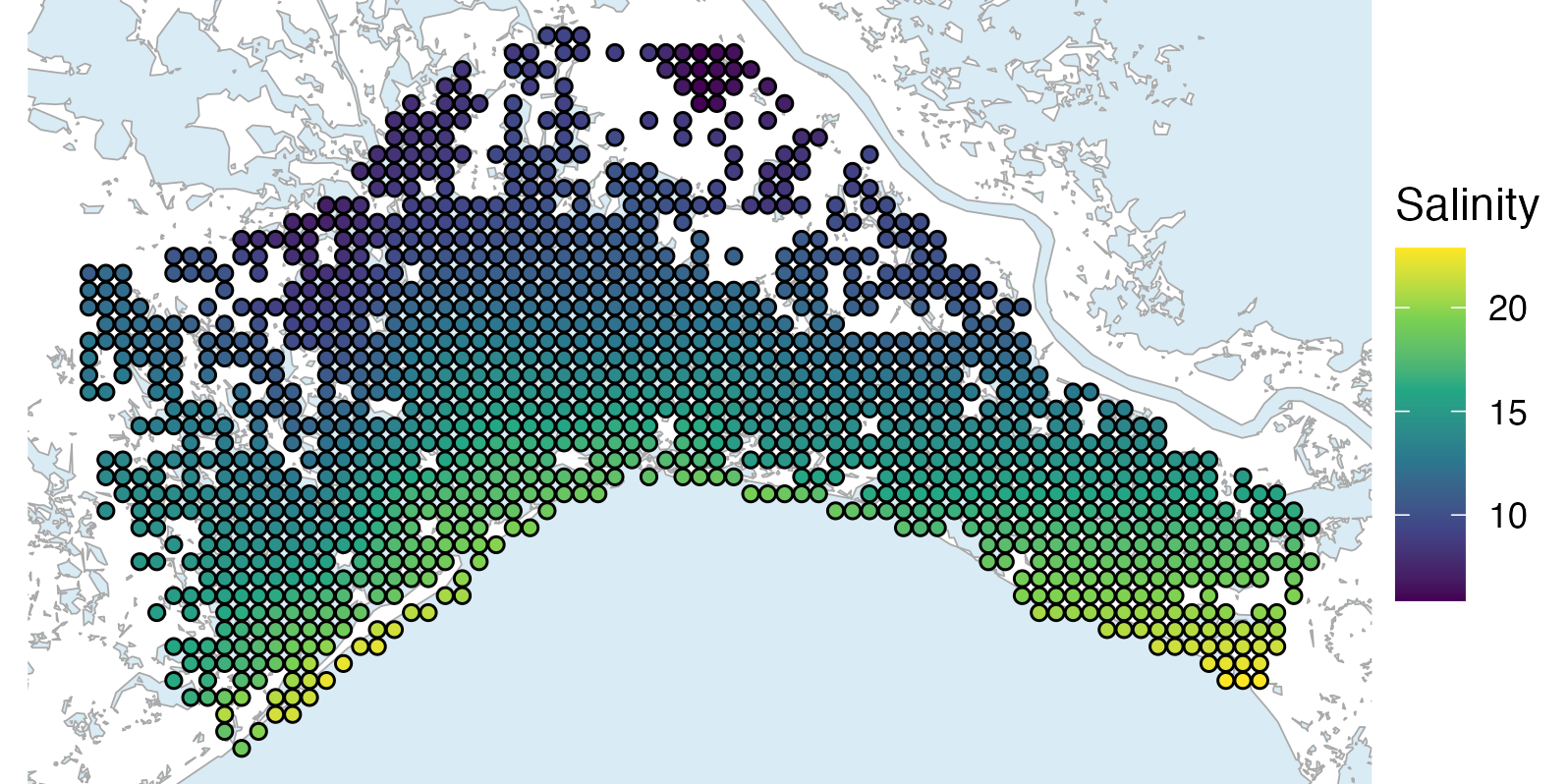} \caption{Salinity averaged over $7$ years ($2011$--$2017$) and over a one kilometer grid}\label{fig:appsalcov}
\end{figure}

Finally, average salinity is a continuous spatial covariate. To compute
the average salinity covariate, first the salinity over a \(10\) meter
grid was averaged over time from Jan \(2011\) to Dec \(2017\). The
average salinity associated with each mesh point is then the
spatial-average of this time-averaged salinity within one kilometer of
the mesh point. Ultimately, the salinity covariate is averaged across
both time and space. Henceforth, the term \textit{averaged salinity}
will refer to this covariate. Supplementary Figure~\ref{fig:appsalcov} shows the
computed averaged salinity.

\hypertarget{appendix-4-salinity-hydrodynamic-modelling-details}{%
\section{Salinity Hydrodynamic Modelling
Details}\label{appendix-4-salinity-hydrodynamic-modelling-details}}

Salinity estimates were from the Delft3D-based hydrodynamic model
described by \citetapp{white2018salinity} and \citetapp{takeshita2021high}.
The model performed historical simulations to estimate salinity
conditions from 2011 to 2017 in the Barataria Basin. It was calibrated
and validated using a collection of field observations, including
salinity, by comparing model output with data from USGS, CRMS, and NOAA.
The model used field measurements (e.g., river discharge and tides) to
impose boundary conditions and atmospheric forces for those specific
years. The Delft3D model uses a 375 m resolution triangular grid in the
majority of our study area, but near the Mississippi River, a small area
of our modeled estimates has a 125 m resolution. Within the model,
salinity estimates are depth averaged. We used publicly available
packages (\texttt{sf}, \texttt{stars}, \texttt{raster}, and \texttt{akima}) for the statistical software \texttt{R}
(version 4.0.0) to generate the estimated salinity averages by
interpolating the daily triangular Delft3D model output into a 375 m
square grid and removing pixels that had no water (i.e., were all land).
We then averaged the salinity estimates in each cell across the entire
available timeframe for estimates from Barataria Basin (January 2011 to
December 2017) to generate a single spatial snapshot of average
salinities across the basin for the multi-year time period.

In the best performing model, recruitment and survival both depended on regression splines of time
whose number of basis functions were selected by comparing model AIC. The optimal model was
found by optimizing the AIC using a method of steepest descent. Supplementary Table~\ref{tab:aics} shows the \(\Delta\)AIC values for 10 models
with the lowest AICs. The only difference among these models was the number of degrees of freedom for the splines, which are displayed in the table. Models with \(\Delta\)AIC \(< 2\)
were taken as part of the best model set that varied only by smoothing hyper-parameters.

\begin{table}[ht]
\centering
\begin{tabular}{rrr}
  \hline
Recruitment $k$ & Survival $k$ & $\Delta$AIC \\ 
  \hline
6 & 5 & 0.0 \\ 
  7 & 5 & 0.2 \\ 
  6 & 4 & 0.5 \\ 
  7 & 4 & 0.7 \\ 
  6 & 3 & 1.5 \\ 
  7 & 3 & 1.7 \\ 
  5 & 5 & 1.8 \\ 
  6 & 7 & 2.2 \\ 
  5 & 6 & 2.3 \\ 
  5 & 3 & 3.2 \\ 
   \hline
\end{tabular}
\caption{$\Delta$AIC (difference between AIC for that model and the AIC for the optimal model) for the best 10 models. The difference between models was the number of basis functions (hyper-parameter $k$) in time-varying population dynamics parameters, survival and recruitment. Only the 7 models with $\Delta$AIC < 2 are included in the best model set.} 
\label{tab:aics}
\end{table}

The AC density surface was selected to depend on regression splines of both
spatial location \((x,y)\) and averaged salinity. A maximum was set on
the number of basis functions of both splines of \(10\) and \(20\) 
respectively. In the case of salinity, AIC clearly selected \(5\)
basis functions as optimal, while for spatial location, the AIC
selected the maximum number of basis functions.

\hypertarget{appendix-5-detection-process}{%
\section{Detection
Process}\label{appendix-5-detection-process}}

\begin{figure}
\includegraphics[width = 0.8\textwidth]{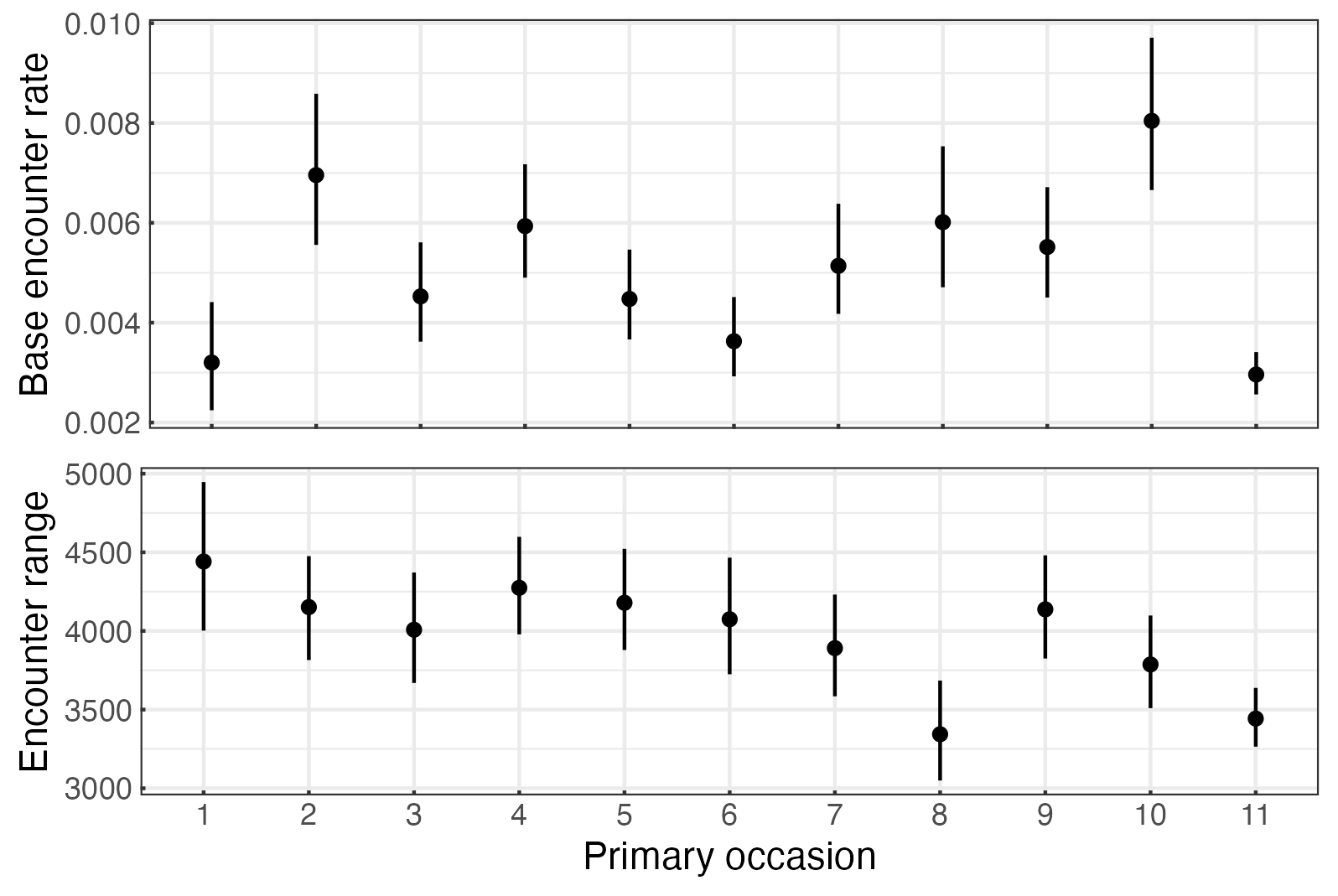}\caption{Mean detection parameter estimates (base encounter rate $\lambda$, encounter range $\sigma$ (m)) and approximate $95\%$ confidence interval over primary occasions (1--11) from $10000$ parametric (model-averaging) bootstrap resamples}\label{fig:det_time_res}
\end{figure}

Base encounter rate varied over time without any clear pattern (Supplementary Figure~\ref{fig:det_time_res}), likely driven primarily by survey conditions.

Encounter range was similar across most primary occasions (Supplementary Figure~\ref{fig:det_time_res}) with a mean of \(3975\)~m (\(3393\)--\(4358\)~m)
with the exception of primaries \(8\) and $11$ where encounter range was lower. Recall that these estimates are not derived from a
regression spline and so any smooth pattern is induced by the data.

\begin{figure}
\includegraphics[width = 0.8\textwidth]{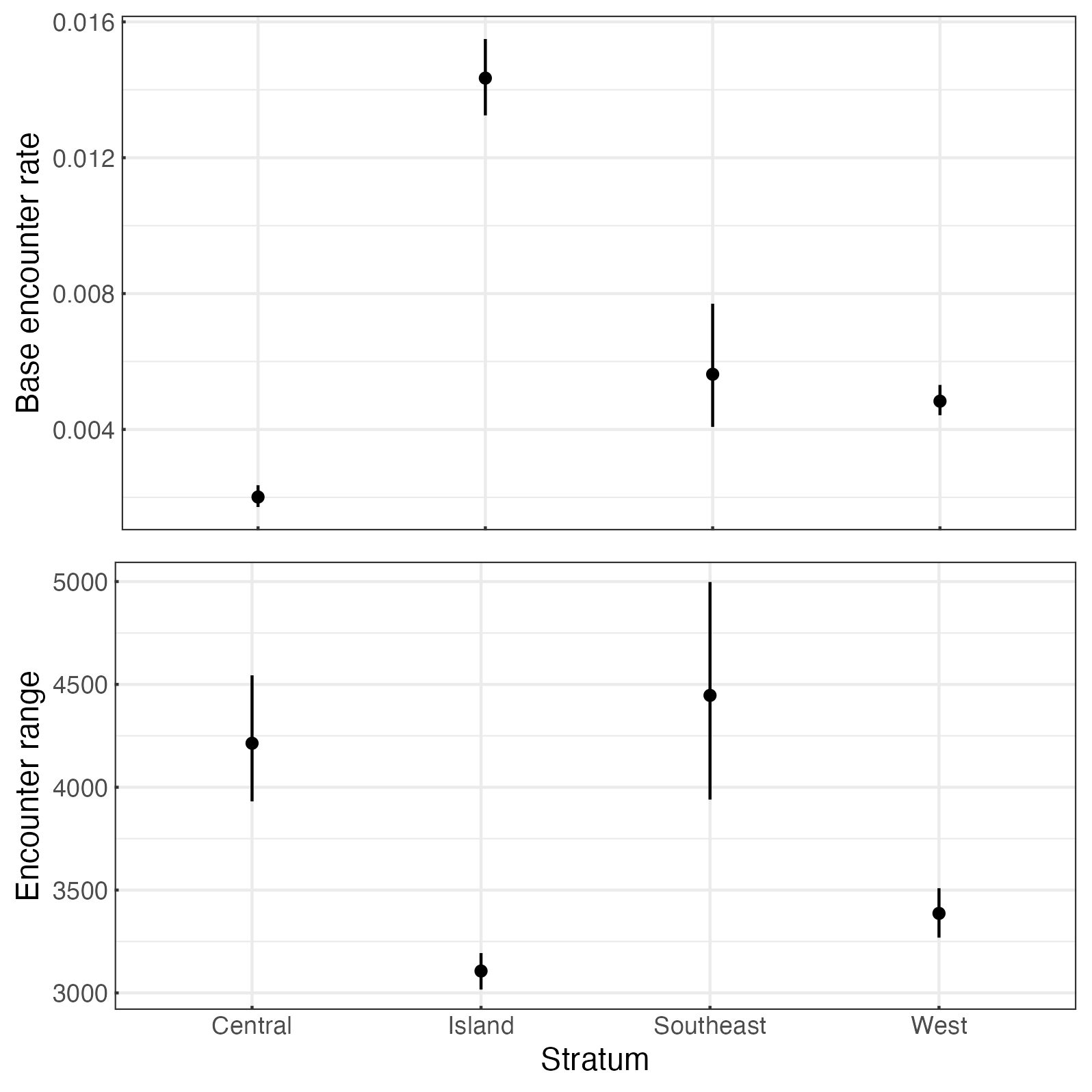}\caption{Mean detection parameter estimates (base encounter rate $\lambda$, encounter range $\sigma$ (m)) and approximate $95\%$ confidence interval for each stratum from $10000$ parametric (model-averaging) bootstrap resamples}\label{fig:detstr}
\end{figure}

Comparing strata, it is best to interpret the base encounter rate and
the encounter range jointly (Supplementary Figure~\ref{fig:detstr}). For the Island
stratum, base encounter rate is much higher than in other strata, while
encounter range is slightly smaller: the reduced encounter range could
simply be due to the relatively small extent of the island stratum
compared to other strata, while the enlarged base encounter rate
suggests an increased encounter rate within the island stratum compared
to surveying that takes place in other strata. The other notable stratum
is the Central stratum. In the Central stratum, the encounter range is
substantially larger while the base encounter rate is much smaller which
suggests that individuals move (or are detectable) over a larger area
around their activity center but that the encounter rate is lower
compared to other strata.

There was no apparent difference between encounter range in the open
and sheltered parts of the study area; however, there was a substantial difference in base encounter rate. Base encounter rate for
surveying in the open was lower than in the sheltered (open: \(0.0020\)
(\(0.0018\)--\(0.0022\)), sheltered: \(0.0066\)
(\(0.0057\)--\(0.0078\))). This could either be due to different
behavior of the dolphins in these two areas leading to disparity in
detection or, and perhaps more likely, is that detection probability is
higher in sheltered areas, particularly sheltered areas in the Island stratum (0.0143).

\hypertarget{appendix-11-bootstrapphi}{%
\section{Bootstrap Survival Estimates}\label{appendix-11-bootstrapphi}}

\begin{figure}[tph]
\includegraphics[width = 0.8\textwidth]{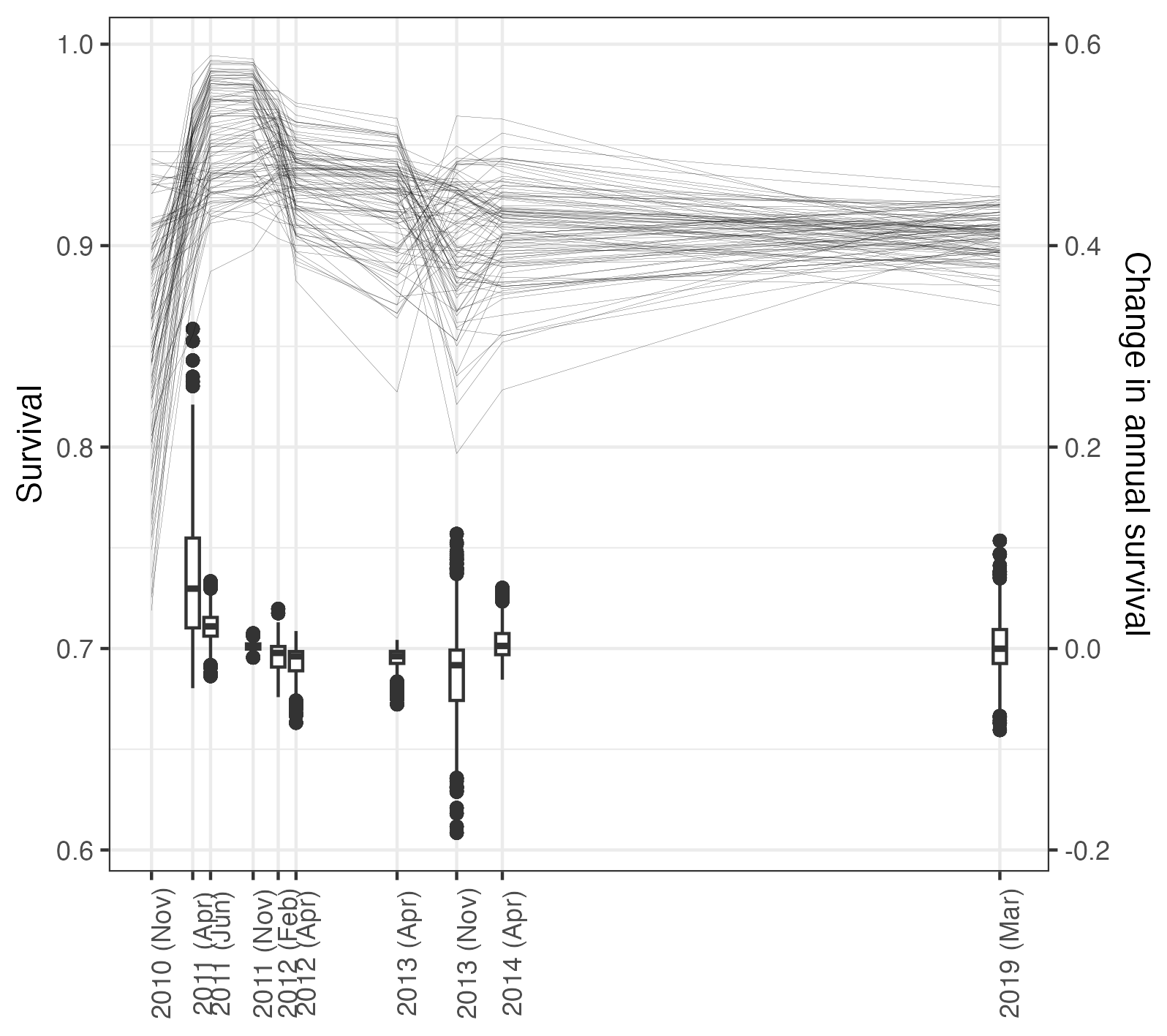} \caption{Random sample of $100$ of the $10000$ bootstrap estimates for survival (black lines) and distribution of the change in annual survival between previous and current primary (boxplots, second y axis).}\label{fig:bootsurv}
\label{fig:bootstrap_survival}
\end{figure}

\begin{table}[ht]
\centering
\begin{tabular}{llll}
  \hline
Date & $\phi$ & LCL & UCL \\ 
  \hline
 09-Nov-10 & 0.86 & 0.73 & 0.95\\
 06-Apr-11 & 0.93 &  0.88  & 0.97\\
 09-Jun-11 &  0.95  & 0.91  & 0.99\\
 09-Nov-11 &  0.95  & 0.92  & 0.99\\
07-Feb-12 &  0.95  & 0.91 &  0.97\\
11-Apr-12  & 0.93 &  0.89 &  0.96\\
09-Apr-13  & 0.92  & 0.87  & 0.96\\
09-Nov-13 &  0.90  & 0.83 &  0.95\\
22-Apr-14 & 0.90  & 0.85  & 0.95\\
14-Mar-19 &  0.91  & 0.88  & 0.93\\
   \hline
\end{tabular} 
\caption{Mean annual survival probability estimates ($\phi$) and approximate 95\% confidence interval (LCL, UCL) over primary occasion from 10000 parametric (model-averaging) bootstrap resamples.} 
\label{tab:phis}
\end{table}
While the estimates of annual survival (\ref{tab:phis})  are the model-averaged estimates from 10000 bootstrap samples, it is useful to inspect the behavior of the individual bootstrap samples in areas of high uncertainty to see if the pattern over time is consistent between samples. We present $100$ randomly selected bootstrap samples of survival in Supplementary Figure \ref{fig:bootstrap_survival}. Since the trend in survival is inconsistent for some bootstrap samples in the first three primaries, we also investigated the proportion of bootstrap estimates that estimate an increase in survival between primary 1 and 2 ($91.3$\%) and primary 2 and 3 ($91.2$\%). This is evidence that survival is increasing from primary 1 to 3, regardless of the uncertainty in the initial survival estimate.

\hypertarget{appendix-1-comparison-w-bayesian}{%
\section{Comparison with Bayesian Methods}\label{appendix-1-comparing-w-Bayesian-methods}}

A previous study by McDonald et al. \cite{mcdonald2017survival} estimated density and survival for this population using a Cormac-Jolly-Seber model with a Bayesian framework implemented in JAGS. We provide a summary of the differences between this previous approach and our improved model, as well as a speed comparison between our model and an updated, more comparable Jolly-Seber model with a Bayesian framework that we implemented in NIMBLE. 

\clearpage

\subsection*{Summary of differences and speed comparison}


\begin{longtable}[]{p{2cm}p{7cm}p{7cm}}
  \hline &
McDonald et al. 2017 & MLE JS Model \\ 
  \hline
Mesh & The mesh included locations of water up to 2km away from the barrier islands. They include 4 strata covariates (east, west, island, and water beyond the islands). The authors acknowledge that ACs from animals detected within the barrier islands may be estimated in the water beyond the islands, but they do not report densities for the this final strata.  & The mesh included locations of water up to 1km away from the barrier islands. They include 4 strata covariates (west, central (corresponding to ``east'' from McDonald et al. 2017), island, and southeast). \\ 
Detection function & Rather than using distance squared as in the half-normal detection function, authors estimated the distance exponent as a third detection function parameter $k$. All detection parameters ($\lambda$, $\sigma$, and $k$) were estimated for each strata. & Half-normal detection function. Both detection parameters were modelled as a linear combination of trap stratum, trap openness, and primary. \\ 
AC density  & AC density is derived from the detection parameter estimates and is estimated for each strata and primary. & AC density is estimated using thin-plate regression spline smooths over a combination of 2-dimensional space ($x$ and $y$) and salinity. \\ 
Abundance &  Abundance is derived from density and the area of each strata, though the areas are not calculated using the same mesh and are instead estimated by another study \citepapp{hornsby_using_2017}. & Abundance is derived from density estimates and the area used to estimate density. \\ 
ACs &  ACs are allowed to move between primaries with a simple movement model that estimated different movement parameters per stratum. Additionally, the authors claim "The spatial capture heterogeneity induced by the SERD model allowed the open (between-primary) portion to infer both permanent and temporary emigration, thereby estimating ‘true’ rather than
‘apparent’ survival," but I do not see evidence for estimating emigration as home range centers can not move outside the study area. & ACs are fixed. \\ 
Survival & Survival was estimated per primary, then averaged over a year of primaries to derive annual survival. & Annual survival was estimated as a rate, which allowed us to account for varying inter-primary periods. We used a thin plate regression spline smooth to model survival as a function of time. \\
Recruitment & Cormack Jolly Seber models do not estimate recruitment. &  We estimated a recruitment rate to account for variable interprimary periods. We used a thin plate regression spline smooth to model recruitment as a function of time. \\ 
Time to \newline fit model & According to authors, it took 9 days to run 3 chains, each with 500 burn in steps and 200 sampling steps on a
single-core 64-bit server. & Average maximum likelihood estimation (MLE) model fit time was between 1 and 3 hours using 6 cores using an Apple M2 Pro chip. We also constructed a Bayesian Jolly-Seber model using NIMBLE with all parameterizations comparable to our MLE model. With the simplest model possible (no covariates, constant parameters) each iteration took approximately 9.3 seconds, which suggests a model run of 30000 iterations would take 3.2 days.\\ 
   \hline
\label{tab:model-comparison}
\end{longtable}


\subsection*{NIMBLE implementation}
  
To efficiently code a Bayesian Jolly-Seber SCR model in NIMBLE, we implemented several custom functions similar to \cite{bischof_estimating_2020}, but ran into some additional challenges and briefly cover our solutions below. We wrote a custom \texttt{nimbleFunction} to implement the forward algorithm to marginalize over the latent animal states similar to the \texttt{nimbleEcology} \texttt{R} function \texttt{dhmm}. However, given the large number of detectors, combined with multiple years, occasions, and unique individuals, we ran into memory allocation issues. To account for that, we included \texttt{setup} code in our distribution function that allowed us to cache most of the data (detector locations and capture histories) within the distribution function and not the model itself. This simplified the model dependencies and reduced memory use, making the setup and compilation steps in NIMBLE much quicker. Due to non-Euclidean distances being used, we precomputed the distance from each detector to a center point on a mesh to approximate the activity centers. Each activity center was given a uniform prior on the mesh, linking to each row and column. For additional efficiency, computing detection probabilities were done only for detectors within a certain maximum distance from the activity center similar to \citepapp{bischof_estimating_2020}. However, one animal in the study was observed at two detectors 29 km apart. As a result, we were only able to limit detection distances to 30 km or less, resulting in still computing the detection probabilities on average of 67\% of the mesh cells.

We followed standard Bayesian SCR approaches and used data augmentation \citepapp{royle2007_analysisofmulti, bischof_estimating_2020}. In this approach, a superpopulation (note that the use of ``superpopulation'' in this Bayesian context is different from the use in the main text) of individuals that were never detected is included as latent nodes. These are sampled, allowing some animals to not exist, and some animals to exist but never be detected. Computation time for each iteration of the MCMC varies depending on the number of augmented individuals that are treated as real but undetected. King et al. \cite{king2016_capturerecap} showed that sampling from the semi-complete likelihood can significantly speed up computation time as opposed to data augmentation. However, given the complexity of the model, it was unclear that that the semi-complete likelihood would be applicable here given the computational challenge of estimating the detection probability of detecting any animal at least once.

Full code implementing the model in NIMBLE is available on Dryad.

\hypertarget{appendix-6-determining-the-region-of-inference}{%
\section{Uncertainty in the Region of
Inference}\label{appendix-6-determining-the-region-of-inference}}

Because abundance is derived using the distinctive individual AC density estimate, is it important to evaluate how uncertainty in the distinctive AC density estimate will contribute to uncertainty in abundance. In most cases, it is not sensible to use the standard deviation to
do this as estimated distinctive AC density has, in theory, a log-normal asymptotic
distribution and so the estimated standard deviation is proportional to
the estimated mean. A popular measure of uncertainty that accounts for
this is the coefficient of variation (CV): the standard deviation
divided by the mean. In this case, however, the CV poorly describes the
uncertainty in the distinctive AC density surface; this is
because the distinctive AC density estimator has a heavily skewed distribution and so
both the mean and standard deviation poorly reflect the center and
spread of the distribution respectively. Thus, in the far north CV is
low because mean and standard deviation are both extremely large due to
the skewness being exacerbated when uncertainty is larger.

\begin{figure}
\includegraphics[width = 0.7\textwidth]{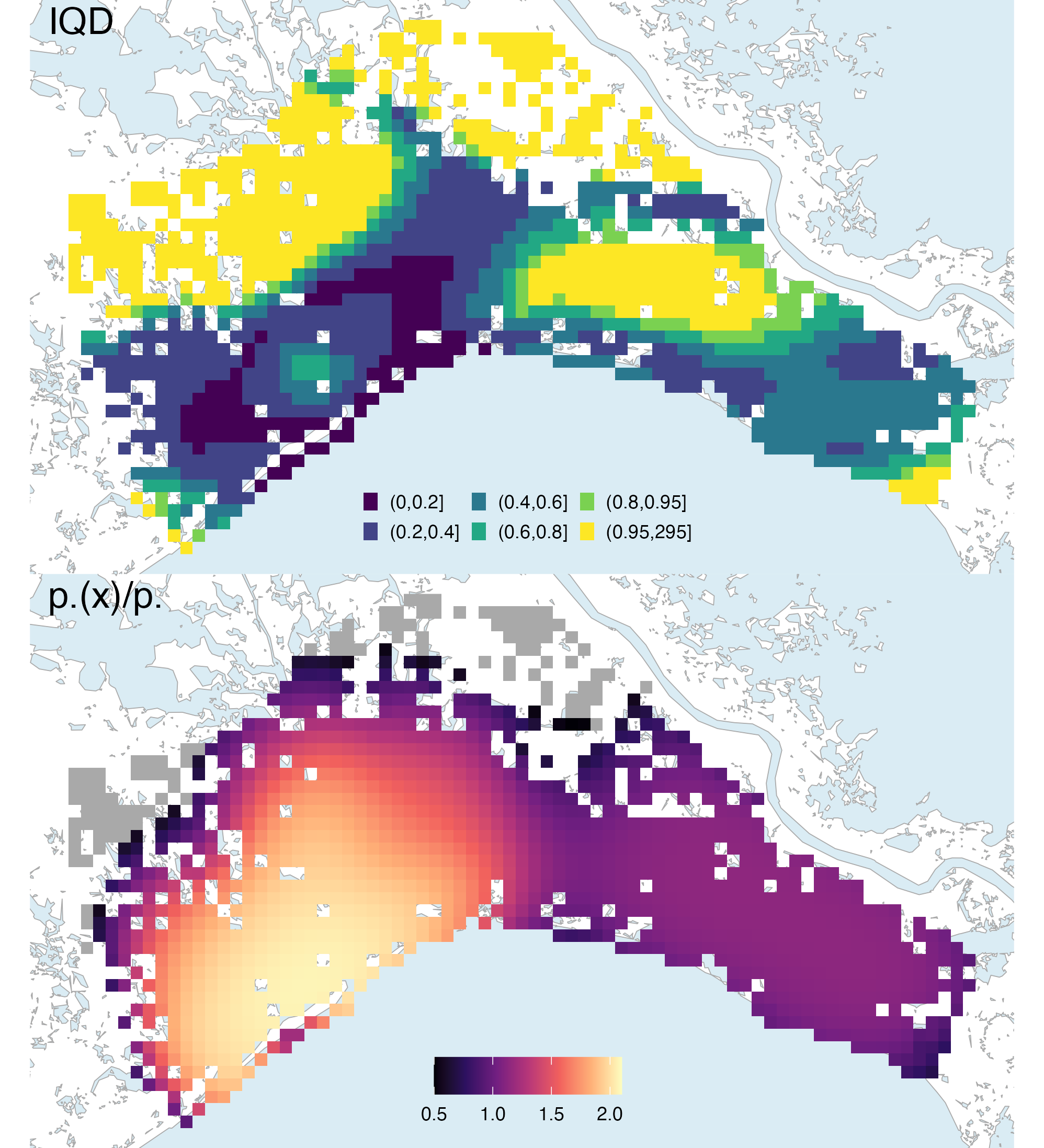} \caption{Interquartile coefficient of dispersion (IQD) for the expected distinctive AC density surface from $10000$ parametric (model-averaging) bootstrap resamples (upper panel) and relative detectability (lower panel). Locations excluded for relative detectability of less than $0.5$ are displayed in grey.}\label{fig:Dcv}
\end{figure}

Thus, as an alternative, we present the interquartile coefficient of
dispersion (IQD), defined as the interquartile range divided by the median. As this is based upon quartiles of the
estimator's distribution, skew and outliers have less effect. We chose a cut-off value of 0.95.

Upon inspection, IQD revealed areas of uncertainty greater than the 0.95 threshold in the northwest, north, and southeastern areas (yellow pixels in Supplementary Figure 7), with some slight variation in areas of uncertainty greater than the cut-off in the southwest region. These areas corresponded with areas with little (southeast) or no survey effort. They also, however, correspond with areas of low distinctive AC density (mean AC density of 0.61). This relatively high uncertainty in distinctive AC density was a combination of both the lack of information from the area and the metrics with which we are estimating uncertainty (i.e., dividing by the mean or median for very small values results in larger uncertainty estimates). Therefore, excluding these locations based on an IQD uncertainty threshold of 0.95 was unlikely to meaningfully reduce uncertainty to the abundance estimate.
Instead, we considered which locations have relatively little information contributing to distinctive AC density estimates using the relative detectability of each potential AC in the mesh. We quantify the relative detectability by calculating the probability of being detected at least once at each give an AC of location $x$ ($p.(x)$) for each location $x$ in the mesh and diving this by the marginalized probability of being detected at least once ($p.$). Locations with a relative detectability ($\frac{p.(x)}{p.}$) less than 0.5 were excluded from inference.

\hypertarget{appendix-7-additional model considerations}{%
\section{Additional Model Checks}\label{appendix-7-additional-model}}

To better understand the variation in survival and recruitment underlying the smooth models, we inspected results from a model (previously fitted during the model selection process) with recruitment and survival dependent on primary as a factor. 
This model was otherwise identical to the best performing model (detection parameters $\lambda$ and $\sigma$ a function of stratum, trap openness, and primary and AC density smoothed over $(x, y)$ space and salinity with 20 and 5 basis functions respectively). By allowing survival and recruitment to vary freely between primaries, rather than being constrained by a smooth as with the AIC-best model, we were able to look for any systematic departures from fits produced by the AIC-best model---such departures could, for example, indicate unaccounted for missing covariates.  Results for this model (which was 9.22 AIC points higher than the best model) are shown in Supplementary Figures \ref{fig:alt_model_params} and \ref{fig:alt_model_ACs}.  Estimates of recruitment and survival from this model were highly variable between primaries but showed no systematic departure from the estimates obtained from the AIC-best model.  Estimates of encounter rate and AC density were nearly identical to the AIC-best model. Together, these results indicate that the variation in survival and recruitment is well described by AIC-best smooth model. 

All models in the candidate set assume each animal's AC is fixed over time.  However, in reality, ACs of animals in the population may move between primary occasion. To check for this, we attempted to fit a new model that allowed ACs to move between primaries via a Markovian process where the expected location of an AC in the next primary follows a Gaussian distribution with a mean of the AC location at the current primary and variance parameter $\sigma_{AC}$ \citepapp{glennie2019open}. Because the true AC locations are unknown, this forms a hidden Markov model. Unfortunately, the model failed to converge, so inference cannot be considered fully reliable. However, even before reaching convergence, estimates of recruitment and survival were nearly identical to those estimated by the best performing model (Supplementary Figure \ref{fig:alt_model_params}). Additionally, detection parameter estimates followed a similar pattern (Supplementary Figure \ref{fig:alt_model_params}). Notably, the estimate of $\sigma_{AC}$ was $246$~m, while estimates of $\sigma$ were $3155 -4497$~m for the moving AC model, which indicates that any movement of ACs between primaries is much less than movement within primaries. Indeed, the sum of the two estimates of $\sigma$ and $\sigma_{AC}$ from the moving AC model are approximately equal to the $\sigma$ estimates from the model averaged bootstraps of a stationary AC model ($3343 - 4441$~m), which reflects the fact that the stationary AC model is effectively estimating a mean AC and a combination of the inter-primary and intra-primary movement with $\sigma$. Given these parameter comparisons, the difficulty of getting models with ACs to converge, and the effective closure of this population, we chose to limit our inference to models with stationary ACs.

\begin{figure}[tph]
    \includegraphics[width = 0.8\textwidth]{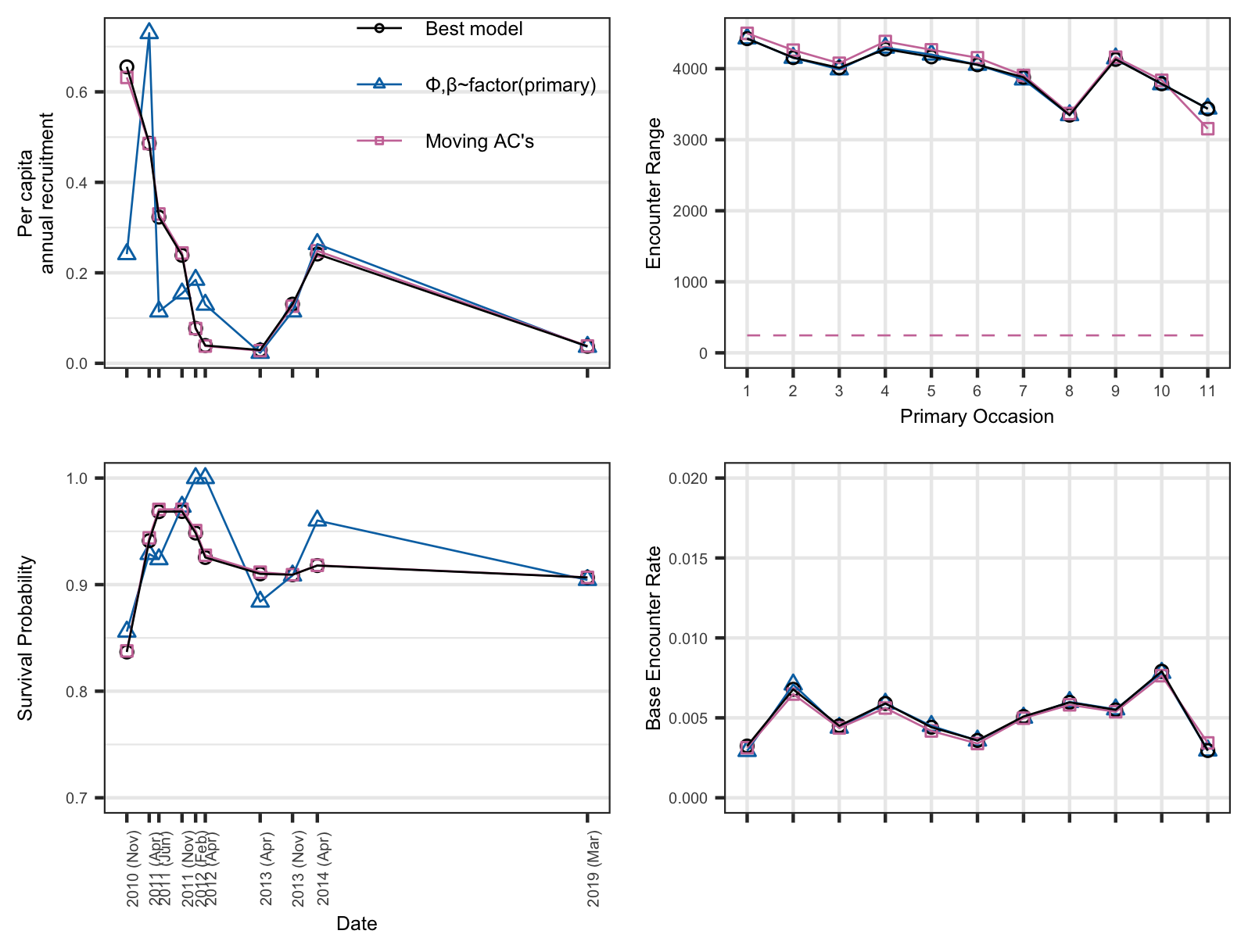}
    \caption{Parameter estimates for per capital annual recruitment (top left), annual survival probability (bottom left), base encounter rate ($\lambda$, top right) and encounter range ($\sigma$, bottom right) for the single best performing model by AIC (black line, round points), a model identical to the best model except survival ($\phi$) and recruitment ($\beta$) are a function of primary as a factor (blue line, triangle points), and a model identical to the best performing model but with ACs that move between primaries according to a Gaussian kernel (pink line, square points). In the top right panel, the pink dotted line displays the estimate of the AC movement parameter ($\sigma_{AC}$), which is the standard deviation of the Gaussian movement kernel. Note that the model with moving ACs failed to converge, so estimates displayed may not reflect the true maximum likelihood estimators.}
    \label{fig:alt_model_params}
\end{figure}

\begin{figure}[tph]
    \includegraphics[width = 0.7\textwidth]{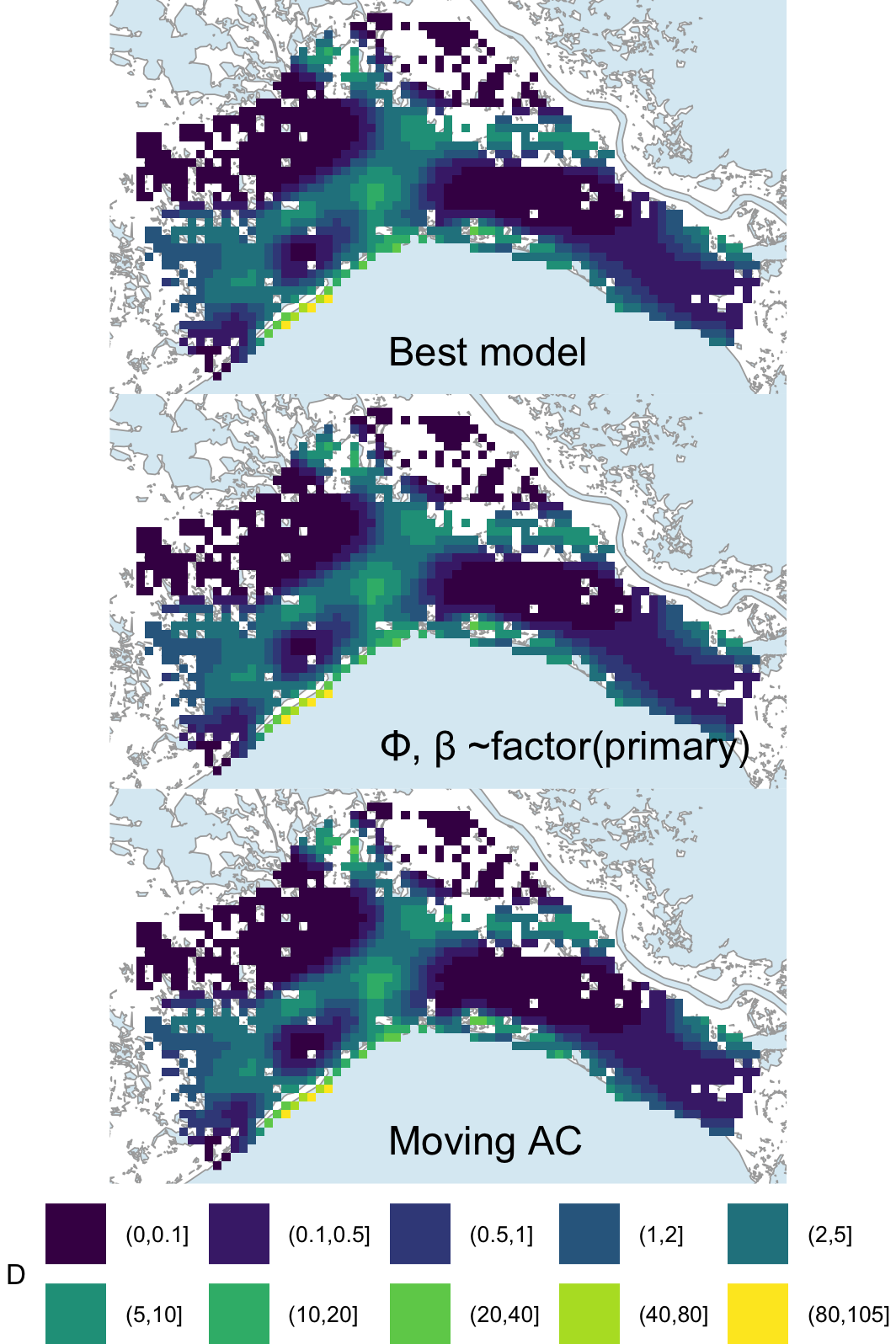}
    \caption{Expected AC density surfaces of distinctive individuals for the single best performing model by AIC (top panel), a model identical to the best model except survival ($\phi$) and recruitment ($\beta$) are a function of primary as a factor (middle panel), and a model identical to the best performing model but with ACs that move between primaries according to a Gaussian kernel (bottom panel). ACs are allowed to move according to a Gaussian kernel, but the model marginalizes over all possible subsequent AC locations, so this density surface represents the average expected AC density of distinctive individuals. Note that the model with moving ACs failed to converge, so density estimates displayed may not reflect the true maximum likelihood estimators.}
   \label{fig:alt_model_ACs}
\end{figure}

\hypertarget{appendix-9-east-west-sight}{%
\section{Surveyed Area in 2019}\label{appendix-9-capthist-summary}}

The survey area was greatly expanded east in 2019. We show the difference in trap coverage from the original study area from 2010 to 2014 and the expanded area from 2019 in Supplementary Figure \ref{fig:counts}. Of the 2091 individuals sighted during the entire study, 1227 were not sighted in 2019, 507 were sighted for the first time in 2019, and 357 were both sighted prior to 2019 and re-sighted in 2019. Of the individuals seen in 2019, very few of them (11) had sightings in both the west and east sides of the study area. We show in Supplementary Figure \ref{fig:counts} that the sighting locations of those individuals that were seen in both the western and expanded eastern areas are most concentrated in the center of the study area. 

\begin{figure}[tph]
    \includegraphics[width = 0.8\textwidth]{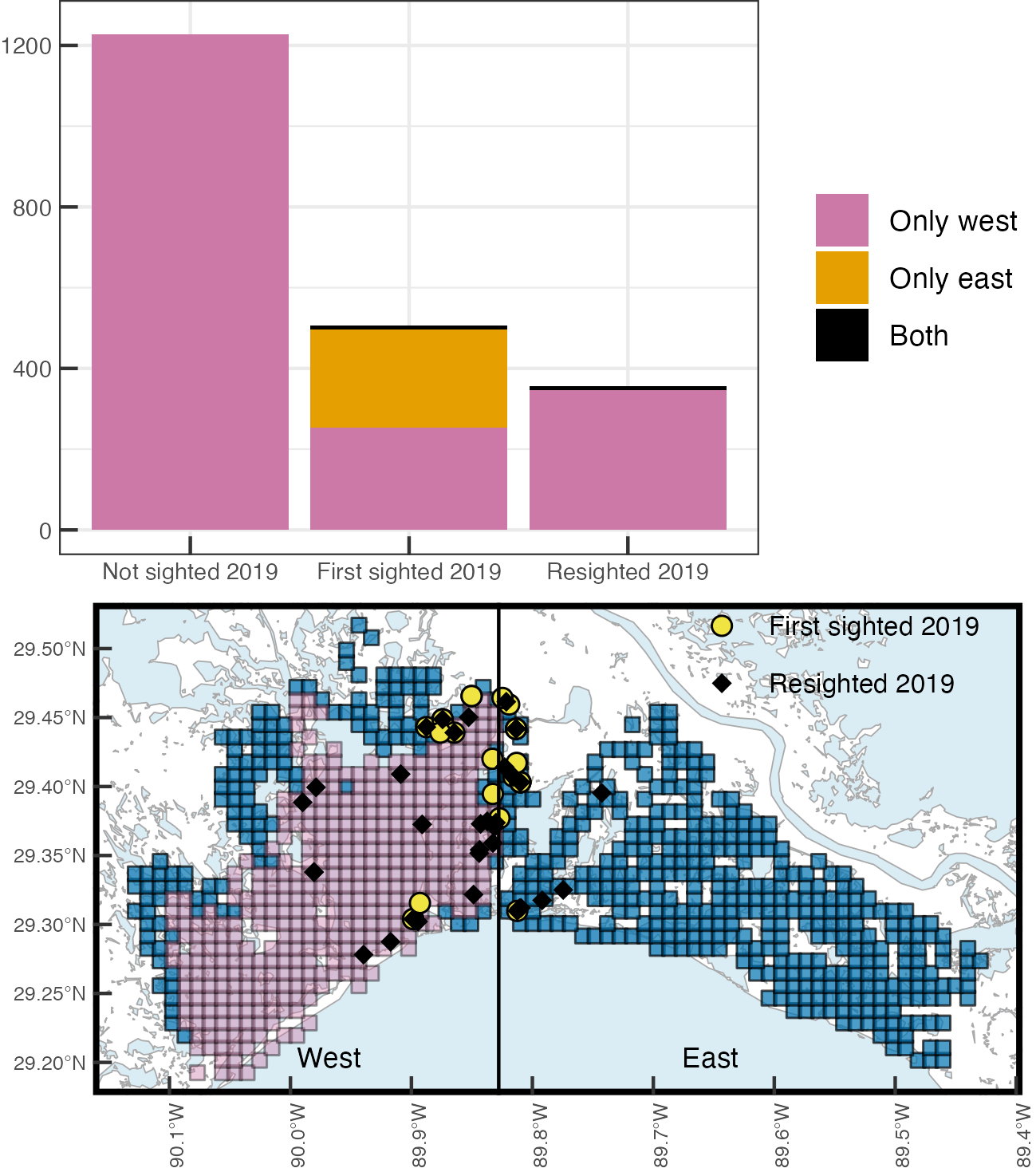}
    \caption{Counts of individuals that were not sighted, first sighted, or resighted in the 2019 surveys over the expanded area (upper panel). Of these individuals, we also display which individuals were only sighted in the western part of the survey area (pink bars, see vertical line in lower panel for east and west distinction), only sighted in the eastern part of the expanded area in 2019 (golden bar), or sighted in both areas (black bar). The lower panel shows the locations of the original traps from 2010 to 2014 (pink squares) and the new traps added in 2019 (dark blue squares) as well as the division we made between the western and eastern study area (black horizontal line). We also include sightings of individuals that were sighted in both the original western area and the eastern area, whether they were first sighted in 2019 (yellow circles) or resighted in 2019 (black diamonds). These points display all sightings of these individuals over the entire study.}
    \label{fig:counts}
\end{figure}

\begin{figure}[tph]
    \includegraphics[width = 0.8\textwidth]{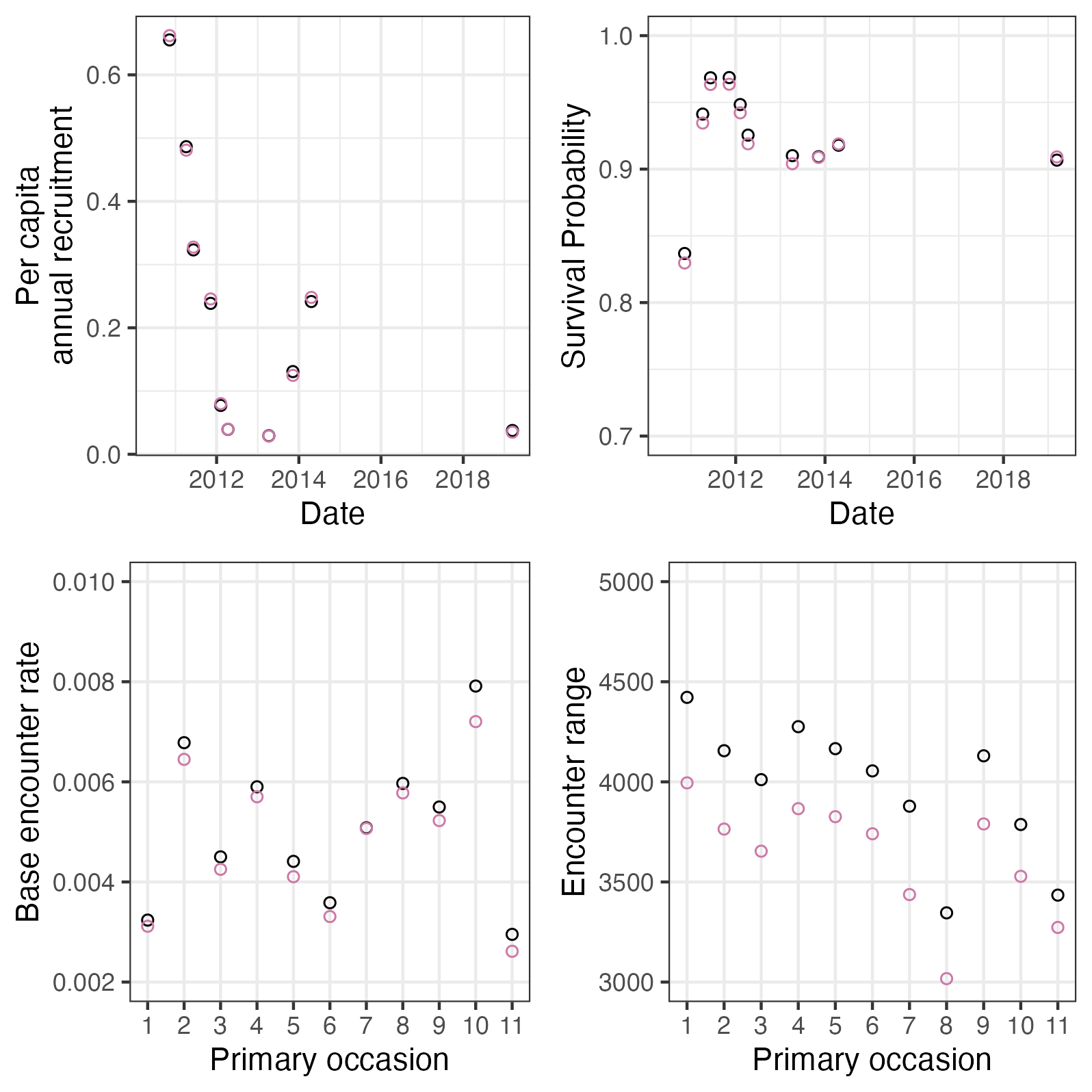}
    \caption{Changes to parameters between the single best performing model fit to the full dataset (black points) or a subset of the capture history that excluded the southeastern strata (pink points).}
    \label{fig:changes}
\end{figure}

To determine if the inclusion of new individuals from this area impacted parameter estimates, we fit the best performing model (detection parameters $\lambda$ and $\sigma$ a function of stratum, trap openness, and primary, recruitment ($\beta$) and survival ($\phi$) smoothed over time with 6 and 3 basis functions respectively, and AC density smoothed over $x y$ space and smoothed over salinity with 20 and 5 basis functions respectively) to a subset of the data that excluded all traps and mesh in the southeast stratum. While this comparison only uses a single model and not a bootstrap-based model average as in the main text, we are still able to determine the impact of including or excluding the southeastern strata on the final inference. When compared with the best performing model fit to the full dataset, the changes to inference are minimal (Supplementary Figure \ref{fig:changes}). The largest difference in per capita recruitment between the full dataset and that excluding the southeast strata was only $-0.007$, while the greatest difference in survival probability was only $0.007$. Maximum changes in detection parameters were also minimal ($0.0007$ and $441.5$m for $\lambda$ and $\sigma$ respectively).

\bibliographystyleapp{apalike}
\bibliographyapp{barb_appendix/appendixreferences}


\end{document}